**Electronic transport and magnetism in the alternating stack of metallic and highly frustrated magnetic layers in $Co_{1/3}NbS_2$**

P. Popčević[1,*], I. Batistić[2], A. Smontara[1], K. Velebit[1,3], J. Jaćimović[4], I. Živković[1,4], N. Tsyrulin[4], J. Piatek[4], H. Berger[4], A. Sidorenko[3], H. Rønnow[4], L. Forró[4,5], N. Barišić[2,3], E. Tutiš[1,†]

[1]*Institute of Physics, Bijenička c. 46, 10000 Zagreb, Croatia*

[2]*Department of Physics, Faculty of Science, University of Zagreb, Bijenička c. 32, 10000 Zagreb, Croatia*

[3]*Institute of Solid State Physics, TU Wien, 1040 Vienna, Austria*

[4]*Laboratory of Physics of Complex Matter, Ecole polytechnique fédérale de Lausanne, 1015 Lausanne, Switzerland*

[5]*Stavropoulos Center for Complex Quantum Matter, University of Notre Dame, Notre Dame, Indiana 46556, USA*

*ppopcevic@ifs.hr

†etutis@ifs.hr

## Abstract

Transition-metal dichalcogenides (TMDs) are layered compounds that support many electronic phases, including various charge density waves, superconducting, and Mott insulating states. Their intercalation with magnetic ions introduces magnetic sublayers, which strongly influence the coupling between host layers, and feature various magnetic states adjustable by external means. $Co_{1/3}NbS_2$ hosts a particularly sensitive magnetic subsystem with the lowest magnetic ordering temperature in the family of magnetically intercalated TMDs, and the only one where the complete suppression of magnetic order under pressure has been recently suggested. By combining the results of several experimental methods, electronic *ab initi*o calculations, and modeling, we develop insights into the mechanisms of electric transport, magnetic ordering, and their interaction in this compound. The elastic neutron scattering is used to directly follow the evolution of the antiferromagnetic order parameter with pressure and temperature. Our results unambiguously disclose the complete suppression of the observed magnetic order around 1.7 GPa. We delve into possible mechanisms of magnetic order suppression under pressure, highlighting the role of magnetic frustrations indicated by magnetic susceptibility measurements and *ab-initio* calculations. Electronic conduction anisotropy is measured in the wide temperature and pressure range. Here we show that the transport in directions along and perpendicular to layers respond differently to the appearance of magnetic ordering or the application of the hydrostatic pressure. We propose the "spin-valve" mechanism where the intercalated Co ions act as spin-selective electrical transport bridges between host layers. The mechanism applies to various magnetic states and can be extended to other magnetically intercalated TMDs.

# I. INTRODUCTION

The research in transition-metal dichalcogenides (TMDs) lives its second golden age, following the first one of the '70s and the '80s of the past century [1,2]. The renewed interest in TMDs is partly driven by fascinating developments in atomically-thin layered systems, triggered by groundbreaking experiments on graphene [3–5], and partly due to discoveries of new electronic states in bulk materials [6–12]. Combining the two seems essential for novel functionalities of quantum devices and future electronics.

The TMDs are quasi-two-dimensional systems with strong in-plane bonding and weak inter-layer coupling. Their electronic system, with effectively reduced dimensionality, is prone to various collective instabilities [1,13]. Moreover, the weak coupling between layers makes TMDs susceptible to intercalation by various atoms and molecules [14,15]. In particular, the intercalations with magnetic atoms provide an exciting playground for studying the interplay between conducting electrons and magnetic lattice degrees of freedom [16–23]. This line of research in TMDs has been only partially explored during their first golden age. Nowadays, it is known that such interplays are essential to produce unconventional electronic states, resulting in elaborated electronic phase diagrams and quantum-critical behaviors observed in many families of materials. Notably, it is precisely in this context that superconducting cuprates [24–29], iron pnictides [30–34], and heavy fermion systems [35–39] are often discussed. The particular advantage of TMDs lies in combining various metallic layers with various magnetic intercalants. Another advantage is the opportunity to fine-tune the coupling between two subsystems by applying relatively modest hydrostatic pressure. In contrast to electronically less anisotropic systems, the magnetic atoms in intercalated TMDs act quite differently in the charge transport in directions parallel and perpendicular to layers. Finally, magnetically intercalated TMDs are known to develop various magnetic phases due to the competition of magnetic couplings of different signs, ranges, and physical origins [14,40–45].

The magnetic order in $Co_{1/3}NbS_2$ appears at a much lower temperature than in other magnetically intercalated TMDs [2]. The ordering temperature $T_N$ (26 K) is approximately six times lower than the Curie-Weiss temperature $\theta$ (157 K), as determined by high-temperature magnetic susceptibility measurements [40]. This fact alone suggests the significant role of magnetic frustration in $Co_{1/3}NbS_2$. The triangular arrangement of magnetic moments within layers may partly be responsible for the frustration, but this feature is not unique to $Co_{1/3}NbS_2$, as it appears in many other magnetically intercalated TMDs. Several research groups have addressed the magnetic structure in Co1/3NbS2 at ambient pressure, and it was identified as the *hexagonal ordering of the first kind* (HOFK) [46]. Uncertainties regarding the exact orientation of spins within layers and a possible non-zero average spin perpendicular to layers [2,40,46,47] have also been addressed to a certain extent [11,48]. The large anomalous Hall effect related to this ferromagnetic component of the magnetic ordering has been reported [11,48]. Recently, it was derived from transport measurements that the ordering temperature decreases under pressure, suggesting a possible complete suppression of magnetic order at elevated pressures [49]. This raises the exciting possibility of the quantum spin liquid emerging at those pressures, embedded between metallic layers and stabilized by magnetic interactions mediated by them.

The mechanism of suppressing the antiferromagnetic (AF) ordering by pressure is also unclear. Several propositions have been made [49]. It was pointed out that Co magnetic moment can get reduced under pressure. This assumption was partly based on some previous experimental findings and supported by a microscopic argument that Co coupling to metallic states increases under pressure. The alternative is that magnetic frustration becomes decisively important under pressure. The results of the present study bring further clarifications related to both aspects.

Here we present a collection of experimental and theoretical results aiming to elucidate the physical circumstances and dominant mechanisms acting in $Co_{1/3}NbS_2$. First, we report the elastic neutron scattering measurements that directly address the pressure dependence of the ordering temperature, $T_N(P)$, documenting the suppression of magnetic order above critical pressure $p_c \approx 1.7$ GPa. Second, the previous in-plane transport measurements [49] are now extended above the $p_c$. Third, we present the first measurements of electrical conduction anisotropy in magnetically intercalated TMDs dependent on temperature and pressure. These measurements, aimed to examine the role of magnetic ions in charge transport in directions along and perpendicular to the $NbS_2$ layers, reveal an unexpected behavior. Fourth, the magnetically ordered state is examined through *ac* and *dc* magnetic susceptibility measurements, confirming the presence of the ferromagnetic component. Finally, we analyze our experimental results in the context of our *ab-initio* electronic calculations, discerning the effects of intercalation on the electronic structure and their implications on electronic transport. The latter analysis profits from the recent ARPES measurements of the electronic structure [12,50,51].

## II. METHODS

Single crystals of $Co_{1/3}NbS_2$ were grown from the vapor phase by iodine transport [2]. The details related to the preparation method are reported elsewhere [40]. The crystal structure of $Co_{1/3}NbS_2$ is derived from one of the parent compound, $2H$-$NbS_2$. Co ions are intercalated at octahedral sites in-between $NbS_2$ layers. They come in a regular triangular planar arrangement, forming the $\sqrt{3}a_0 \times \sqrt{3}a_0$ superstructure, $a_0$ being the lattice constant of the hexagonal unit cell of the parent compound ($a_0$ = 0.331 nm in $2H$-$NbS_2$ [52]) $Co_{1/3}NbS_2$ crystalizes in the hexagonal unit cell (space group $P6_322$ and Pearson symbol hP20) [40]. The crystal axis *c* is perpendicular to layers, whereas the axes *a* and *b* run along layers.

Unit cell parameters of single crystal $Co_{1/3}NbS_2$ were determined at room temperature using Oxford Diffraction Xcalibur Nova R diffractometer with microfocus Cu tube ($K\alpha$ line at 1.54184 Å). The data reduction and the calculation of unit cell parameters were made using the CrysAlis PRO program package [53]. Two large plate-like single crystals (crystal 1 $(0.50 \times 0.45 \times 0.03\ mm^3)$ and crystal 2 $(0.40 \times 0.20 \times 0.02\ mm^3)$) were selected for the determination of the unit cell. For both crystals, we collected the diffraction data to a completeness of around 90 %. The unit cell parameters were calculated from 217 and 234 reflections for crystals 1 and 2, respectively. It was determined that, at room temperature, $a$ = 0.576 nm and $c$ = 1.186 nm. The comparison with the parent compound $2H$-$NbS_2$ shows that the intercalation weakly affects the $NbS_2$ planes and their separation [54].

The samples for electrical resistivity measurements were cut into rectangular forms of sizes 1×0.2×0.03 $mm^3$ and 0.8×0.8×0.06 $mm^3$, with the smallest dimension being perpendicular to $NbS_2$ planes. Gold wires for transport measurements were attached to the crystals using the DuPont silver paste 6838 and cured in a vacuum for 10 min at 200 °C. Electrical resistivity along the *c*-axis ($\rho_c$) was measured on the larger crystal. For this measurement, the current contacts were painted in a circular form on 20% of the two opposite largest surfaces. Dot-like voltage contacts were painted inside those circles. Much care was invested in the precise alignment of these contacts. Still, the error regarding the absolute value of $\rho_c$ is estimated to be up to 50%. The results for $\rho_c$ in 2H-$NbS_2$ and $Co_{1/3}NbS_2$ at ambient pressure were confirmed using focused ion beam (FIB) sample fabrication [55]. Electrical resistivity under pressure up to 2.5 GPa was measured using the self-clamped piston-cylinder pressure cell. The pressure was monitored *in situ* by measuring the resistance of a standard manganin pressure gauge. The pressure medium used was Daphne oil 7373. The reliability and high precision of the used high-pressure experimental setup were already confirmed in other investigations [56].

The neutron scattering measurements under pressure were performed using the triple-axis spectrometer at Institut Laue–Langevin, Grenoble, using the in-house-made pressure cell [57]. The dc magnetic susceptibility was measured using the SQUID magnetometry in the temperature range 2 K – 300 K, and the magnetic field applied parallel or perpendicular to the *c*-axis. We measured the isothermal magnetization curves at different temperatures after cooling in zero field (zero-field-cooling, ZFC). The non-SQUID CryoBIND ac susceptibility system [58] was used to measure the ac susceptibility, with the magnetic field (10 Oe rms) applied within the *ab*-plane of the crystal sample.

*Ab-initio* calculations were made using the Quantum ESPRESSO package [59], with ultrasoft pseudopotentials from Pslibrary [60]. The kinetic energy cut-off for wave functions was 70 Ry, whereas the kinetic energy cut-off for charge density and potential was 600.0 Ry. We have used the PBE exchange energy functional [61] and the Marzari-Vanderbilt smearing [62] of the Fermi surface of 0.005 Ry. The Brillouin-zone sampling used in self-consistent calculations for $2H$-$NbS_2$ was $19 \times 19 \times 5$ k-points (with no shift) and $10 \times 6 \times 5$ k-points (with no shift) for $Co_{1/3}NbS_2$. The density of states (DOS) and the Fermi surface were calculated using a denser k-point mesh. On-site Coulomb interaction on Co ions was taken into account within the DFT+$U$ approach proposed by Cococcioni and de Gironcoli [63]. The value of $U$ was obtained from DFT using the linear-response theory [64]. To compare energies of different magnetic ground states, larger kinetic/charge density and potential cut-offs of 90.0/1100.0 Ry were used to obtain consistent results. The crystal structures are relaxed within calculations unless the contrary is stated.

## III. EXPERIMENTAL RESULTS

### A. Effect of intercalation on electrical resistivity

Fig. 1 shows the temperature dependence of electrical resistivity at ambient pressure in $Co_{1/3}NbS_2$ and $2H$-$NbS_2$ (the parent compound), measured in directions parallel ($||ab, \rho_{ab}$) and perpendicular ($||c, \rho_c$) to $NbS_2$ layers. The resistivity measured in the direction parallel to layers shows metallic temperature dependence in both compounds. $2H$-$NbS_2$ becomes superconducting at 6.1 K (with extrapolated residual resistivity of 1.35 $\mu\Omega$cm and residual-resistivity ratio (RRR) of 80), whereas the intercalated compound exhibits relatively high residual resistivity whose origin is not completely understood [2,49]. A possible cause is the domain structure of intercalated ions, which is known to be present in intercalated TMD compounds [65].

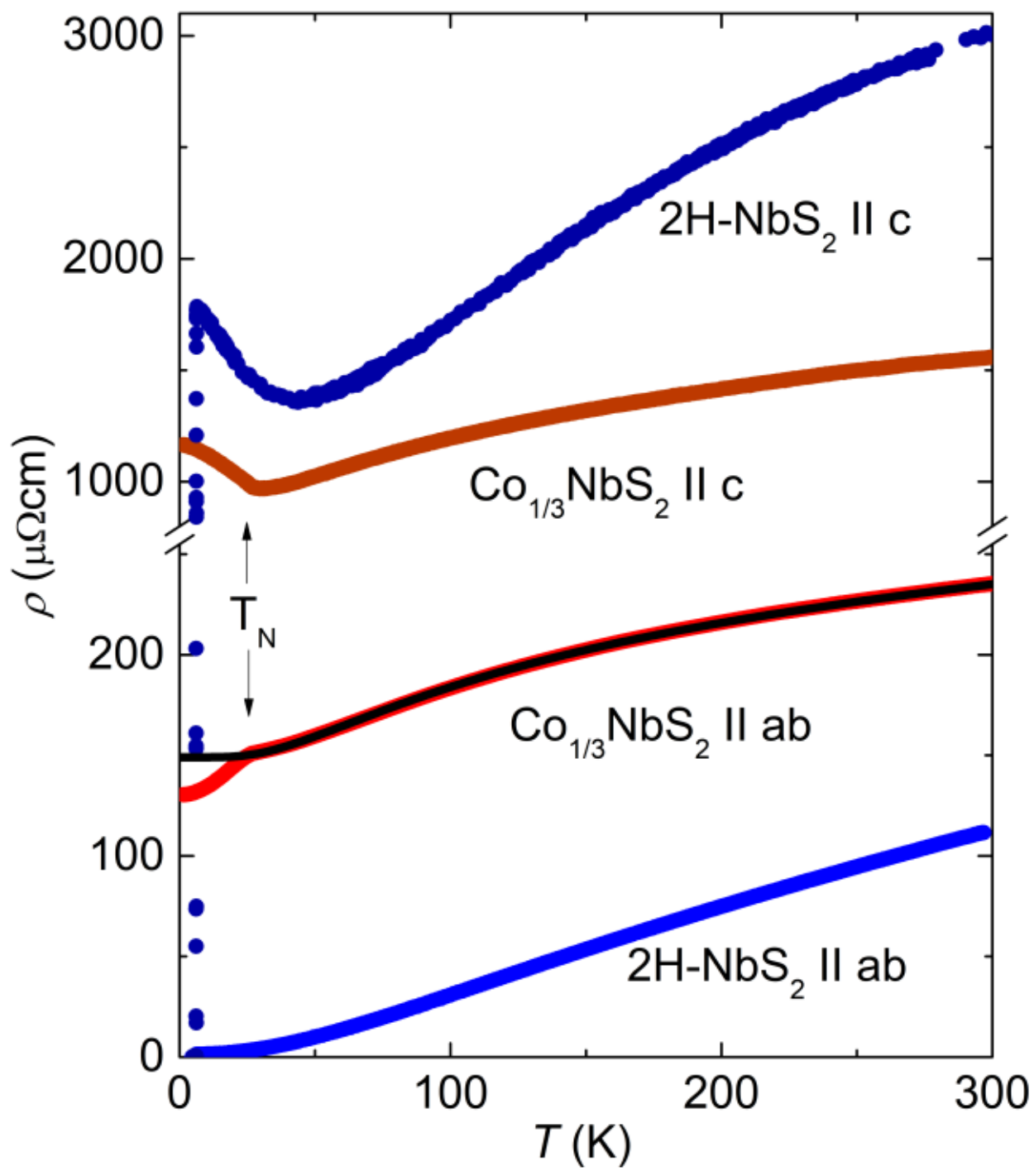

Fig. 1. The electrical resistivity of $Co_{1/3}NbS_2$ and its parent compound $2H$-$NbS_2$ measured in the direction parallel ($||ab$) to $NbS_2$ layers and along the $c$-axis ($||c$), perpendicular to layers. The black line is fit to Eq. (1). Note that $c$-axis resistivity is significantly lower in $Co_{1/3}NbS_2$ than in $2H$-$NbS_2$. The opposite is the case for the in-plane conductivity, implying much lower resistance anisotropy in the intercalated system. Note that the magnetic ordering in $Co_{1/3}NbS_2$, occurring below $T_N$=26 K, has opposite effects on electrical resistivity in directions parallel and perpendicular to layers.

Regarding the electrical resistivity in the direction perpendicular to layers, $\rho_c$, it is instructive to start from $2H$-$NbS_2$ where $\rho_c$ is much higher than $\rho_{ab}$. The anisotropy, $\rho_c/\rho_{ab}$, already substantial at room temperature, $(\rho_c/\rho_{ab})_{RT} = 27$, rises to a much larger value at low temperatures, $(\rho_c/\rho_{ab})_{6K} = 1300$. This strong temperature dependence of anisotropy in $2H$-$NbS_2$ is primarily the consequence of much bigger residual resistance in $\rho_c$ than in $\rho_{ab}$. This property was recently attributed to the $1T$-$NbS_2$ inclusion layers [66].

Turning to $Co_{1/3}NbS_2$, the $c$-axis resistivity at room temperature is two times smaller in the intercalated compound than in the parent compound. The difference in the interlayer distance can be partly responsible, only slightly smaller in intercalated than in the parent compound [67]. The hybridization between Co-orbitals and $NbS_2$ layers probably contributes by providing an additional electronic conduction channel. We further discuss this in Section IV.C and Appendices A and C, where the material's electronic structure is examined. Fig. 1 also shows that the magnetic ordering in $Co_{1/3}NbS_2$ is accompanied by the upturn in $\rho_c(T)$ at Néel temperature, followed by a monotonic rise upon further cooling. This increase of resistivity upon spin-ordering is counter-intuitive to some degree. Generally, one expects the electronic scattering and electrical resistivity to decrease upon reducing the spin disorder. On the other hand, the expected downturn in resistivity in the magnetically ordered state is found for the in-plain resistivity component, $\rho_{ab}(T)$.

## B. Effect of pressure on magnetic ordering and electronic transport

The elastic neutron diffraction measurements provide a way to verify directly the hydrostatic pressure's effect on magnetic ordering. Here we use the single crystal of $Co_{1/3}NbS_2$ from the same batch as those employed for transport experiments. We identify magnetic peaks that match the magnetic structure determined earlier [46]. The temperature dependence of the intensity of the reflection at [0.5,0.5,0] at ambient pressure, shown in Fig. 2 (a), reveals the long-range magnetic order setting in at 26 K. The wave vector, corresponding to the unit cell doubling in the magnetically ordered state, is also the M-point of the first Brillouin zone of the high-temperature phase. Fig. 2 (b) shows the variation of scattering intensity around [0.5,0.5,0] at 10 K under pressure, up to the complete disappearance of the signal around 1.7 GPa. The data in Fig 2. unambiguously demonstrate that the primary claim about the phase diagram of $Co_{1/3}NbS_2$, previously inferred from transport measurements, is correct [49].

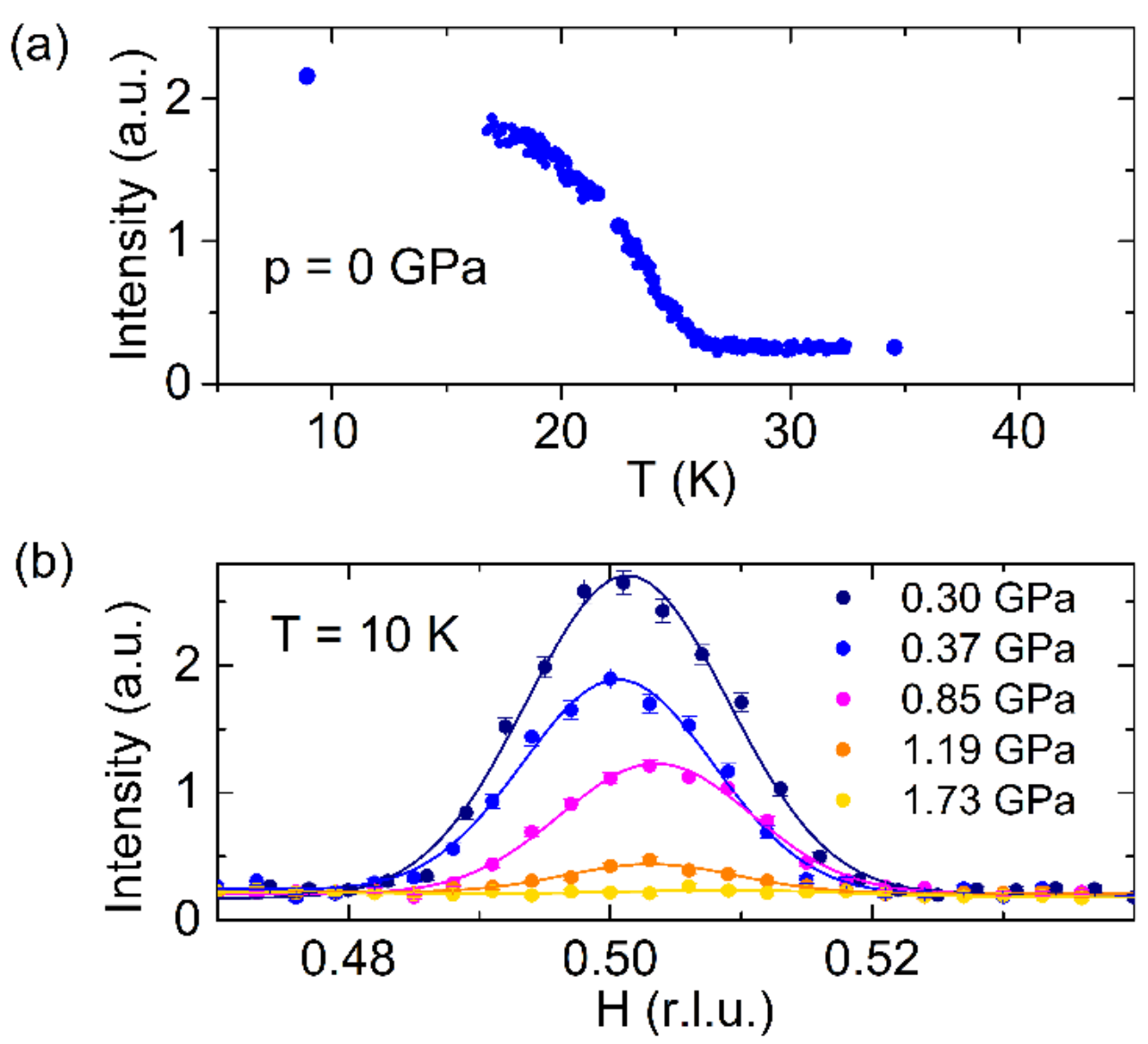

Fig. 2. (a) Temperature dependence of the intensity of the [0.5,0.5,0] magnetic peak at ambient pressure. (b) Pressure dependence of the intensity of the [0.5,0.5,0] magnetic peak at 10 K.

The evolution of in-plane resistivity $\rho_{ab}(T)$ under pressure up to 1.6 GPa has already been described in Ref. [49]. There, the magnetic ordering temperature was related to the minimum in $(d\rho_{ab}/dT)$. Here we extend the transport measurements under pressure by measuring resistivity in the direction perpendicular to layers, $\rho_c$, and by expanding the pressure range well above the critical value.

Fig. 3 shows that in the pressure range between 1.7 GPa and 2.36 GPa $\rho_{ab}(T)$ becomes progressively more featureless below 30 K and acquires the ordinary metallic temperature dependence at higher pressures. In the pressure range between 2.0 and 2.4 GPa, we have extended our measurements to 60 mK, motivated by previous findings of superconductivity in the vicinity of the magnetically ordered phase [7,30,68,69]. However, no superconductivity was observed in the pressure range just quoted or at any other point in the investigated pressure-temperature phase diagram.

Interestingly, the changes in $\rho_{ab}$ and $\rho_c$ induced by pressure are opposite in sign in the wide high-temperature range between 50 and 300 K. It can also be noted that the upturn in $\rho_c(T)$, related to magnetic ordering at ambient pressure, diminishes upon raising the pressure. The temperature of minimum steadily declines under pressure. The minimum persists to pressures slightly above 1.7 GPa, and $\rho_c(T)$ maintains multiple inflections in the low-temperature region even at higher pressure.

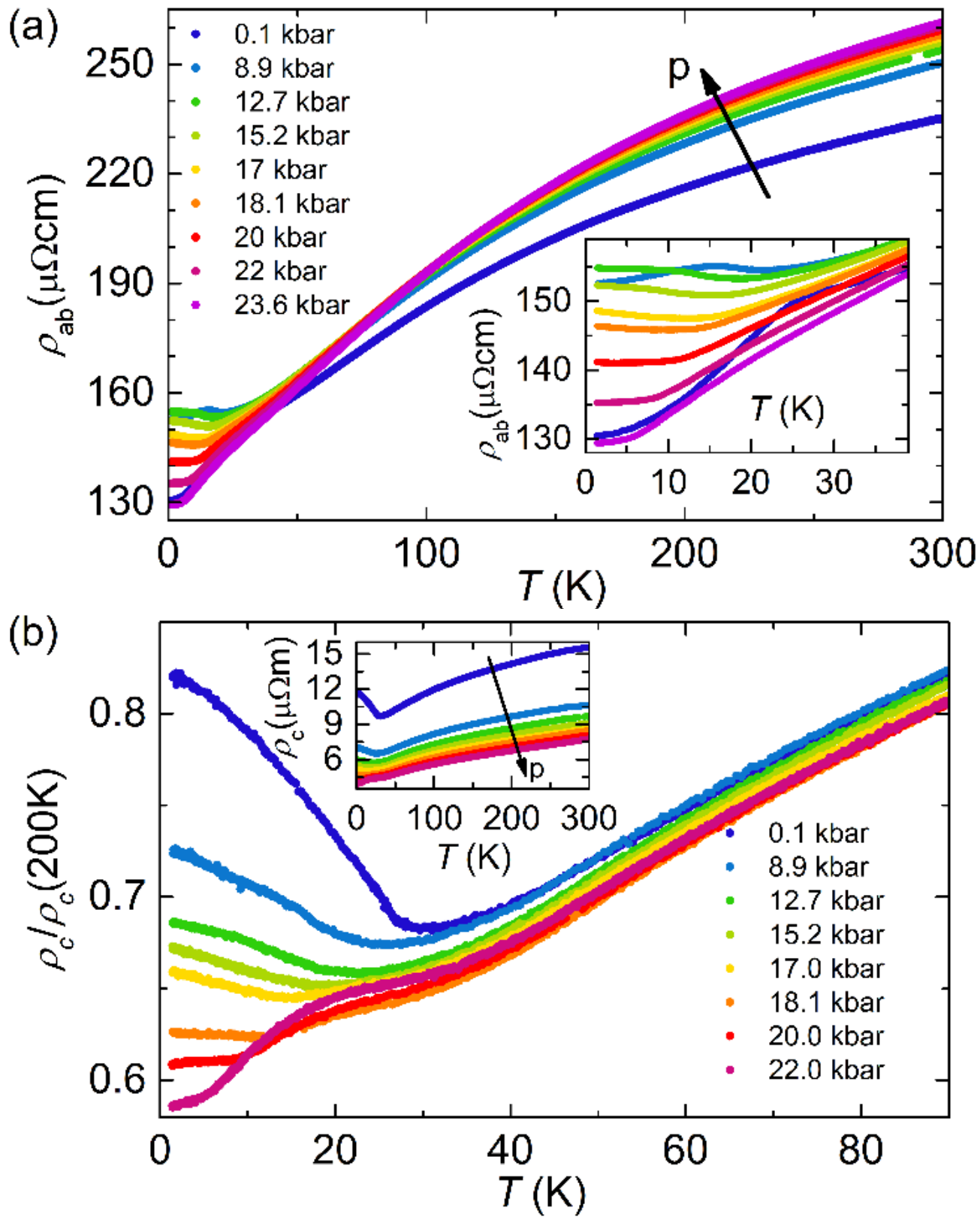

Fig. 3. The electrical resistivity of $Co_{1/3}NbS_2$ measured (a) in the *ab*-plane ($\rho_{ab}$) and (b) along the *c*-axis ($\rho_c$) measured at different pressures in the 1.5 – 300 K temperature range.

Based on these experimental results, we present the extended *P-T* phase diagram of $Co_{1/3}NbS_2$ in Fig. 4.

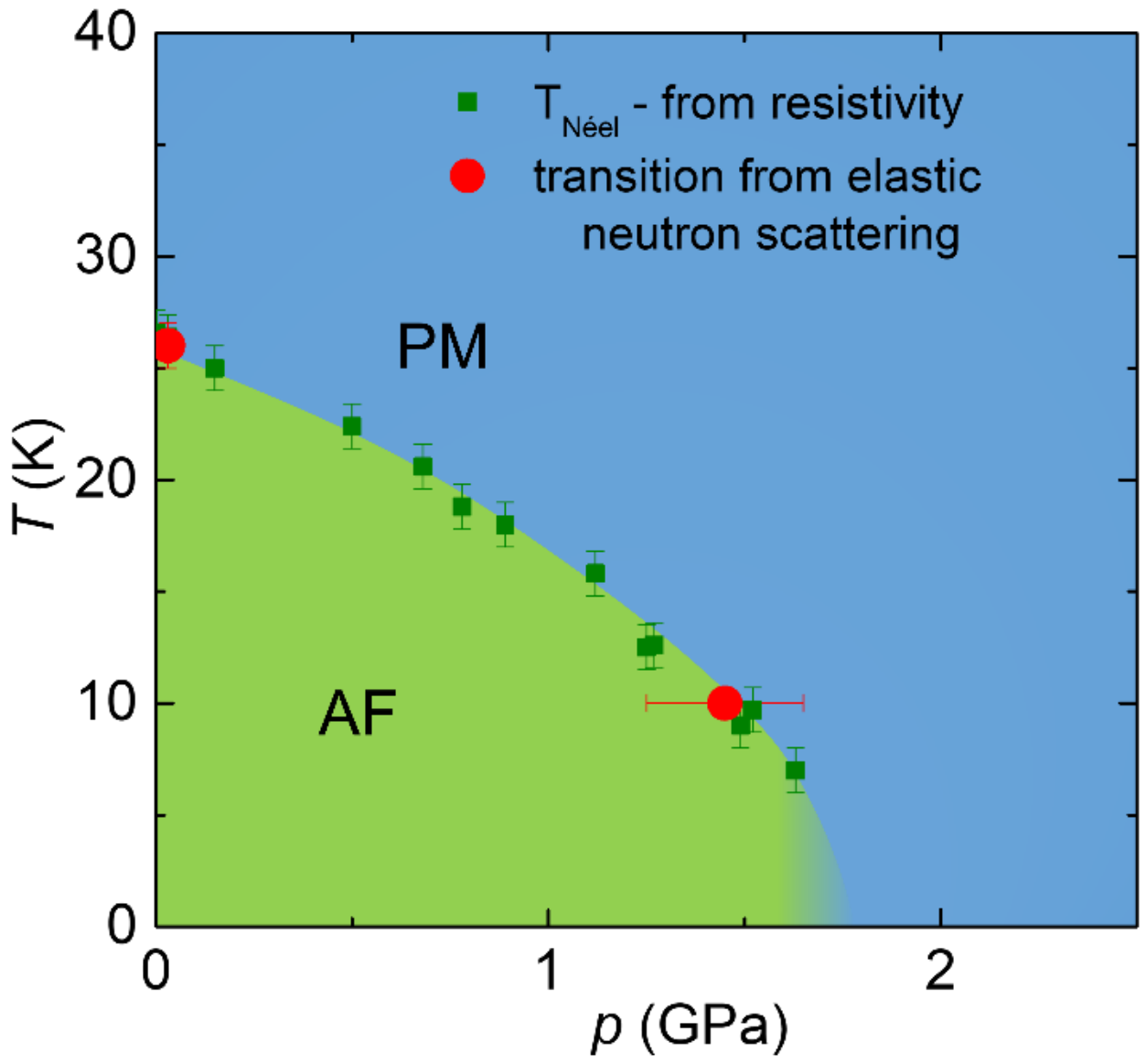

Fig. 4. Phase diagram of $Co_{1/3}NbS_2$ under pressure. The green portion represents the antiferromagnetic (AF) phase while blue stands for the paramagnetic (PM) phase. The green squares represent the inflection point of the electrical resistivity curve $\rho_{ab}(T)$; the red circles mark the phase transition obtained from elastic neutron scattering.

## C. Magnetic susceptibility and frustration

Fig. 5 (a) shows our results for the magnetic susceptibility of $Co_{1/3}NbS_2$ measured below room temperature. The two curves correspond to configurations with the magnetic field of 1T oriented parallel and perpendicular to the *c*-axis. In both cases, the susceptibility follows, from the highest measured temperature to below 100 K, the Curie-Weiss law, $\chi(T) = \chi_0 + \mathrm{C}/(\mathrm{T}-\theta)$. As usual, $\chi_0$ represents temperature-independent contribution resulting from diamagnetism and Pauli paramagnetism, $C$ stands for the Curie constant, and $\theta$ is the Curie-Weiss temperature [70]. For the system of identical magnetic ions, the Curie constant is given by their concentration $n$ and the square of their magnetic moment $\langle\vec{\mu}^2\rangle$, $C = n\langle\vec{\mu}^2\rangle/3\mathrm{k_B}$. Using the concentration of Co ions in $Co_{1/3}NbS_2$ for $n$, our measurements yield $\sqrt{\langle\vec{\mu}^2\rangle} = (3.17 \pm 0.03)\mu_B$, $\theta = (-170 \pm 5)K$ and $\chi_0 = (0.6 \pm 0.5) \times 10^{-4}$ emu/mol Oe for the magnetic field applied parallel to *ab*-plane, and $\sqrt{\langle\vec{\mu}^2\rangle} = (3.13 \pm 0.05)\mu_B$, $\theta = (-160 \pm 5)K$ and $\chi_0 = (1.5 \pm 0.5) \times 10^{-4}$ emu/mol Oe applied along the *c*-axis. These values correspond rather well to those reported earlier [11,40,47].

The Curie-Weiss temperature is much higher than the magnetic ordering temperature, with the "factor of frustration", $|\theta|\ /T_N$, larger than 6, corroborating the assumption of strong magnetic frustration [71]. The magnetic frustration in $Co_{1/3}NbS_2$ may partly come from the triangular arrangement of Co within Co sub-layers [32]. On the other hand, the competing magnetic couplings reaching beyond the nearest neighbors may also contribute to the large factor of frustration.

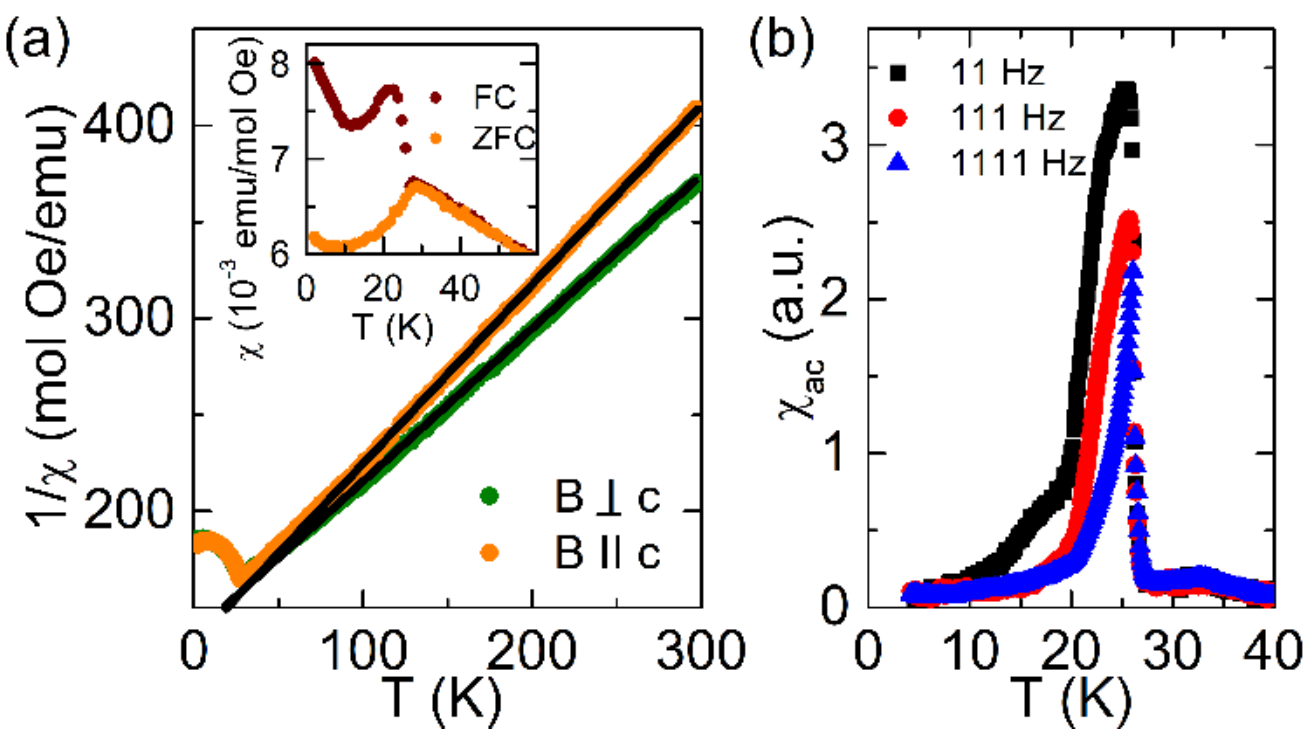

Fig. 5. (a) The magnetic susceptibility measured in the zero-field-cooled (ZFC) regime. The measurements were conducted with a 1 T magnetic field oriented in the *ab*-plane (olive circles) and along the *c*-axis (orange circles). Black lines are fits to the Curie-Weiss law, as explained in the text. Inset: Magnetic susceptibility in field-cooled (FC, dark red circles) and ZFC (orange circles) regimes measured using 0.1 T magnetic field oriented along the *c*-axis, (b) ac magnetic susceptibility measured at frequencies of 11, 111 and 1111 Hz in the field of 1 mT rms.

The inset in Fig. 5 (a) shows the difference in the magnetic susceptibility measured in FC vs. ZFC regimes with the magnetic field pointing along the *c*-axis [72]. In Fig. 5 (b), we also present *ac*-magnetic susceptibility measurements that show transition at 26 K accompanied by a substantial signal below the transition temperature. The latter indicates the ferromagnetic character of the ordered state. This we attribute to canting of magnetic moments along the *c*-axis, simultaneously occurring with their antiferromagnetic ordering. Consequently, the appearance of the ferromagnetic component also provides a simple explanation for the observed large anomalous Hall effect [11], which seems unique to Co-intercalated TMDs [11,73,74]. It is worth emphasizing that the ferromagnetic canting of AF-ordered magnetic moments usually results from antisymmetric exchange (Dzyaloshinskii-Moriya (D-M)) contribution to the magnetic exchange interaction [75]. That points to super-exchange [76] as the relevant interaction mechanism between Co magnetic moments in $Co_{1/3}NbS_2$. Notably, the D-M interaction is held responsible for the helical magnetic order in the iso-structural sister-compound $Cr_{1/3}NbS_2$ [44,45,77].

## IV. DISCUSSION AND THEORETICAL CONSIDERATIONS

### A. Understanding electronic transport

Here we start with the temperature dependence of the in-plane resistivity, $\rho_{ab}(T)$, which shows concave curvature between 50 K and 300 K. Similar concave behavior in electrical resistivity has been observed in several other systems, such as A15 superconductors [78,79], some spinel and skutterudite compounds [80–82], the filled cage compounds [83], and the Kondo lattice systems [84]. The common feature appearing in the explanations of such behavior is "an effective dynamic disorder" that scatters electrons and grows upon heating but saturates above some energy/temperature scale. The concave behavior in resistivity is usually parametrized by extending the usual Bloch-Grüneisen relation by a phenomenological term with thermally activated behavior [78],

$$\rho_{ab}(T) = \rho_0 + \rho_1 \left(\frac{T}{\theta_D}\right)^n \int_0^{\theta_D/T} dx \frac{x^n e^x}{(e^x-1)^2} + \rho_2 e^{-T_0/T}, \tag{1}$$

where $\theta_D$ and $T_0$ represent the Debye temperature and characteristic temperature for thermal activation of disorder, respectively. The relation (1) fits the temperature dependence of the in-plane resistivity observed in Co1/3NbS2 at ambient pressure very well. The fitting provides $\theta_D$ = (400 ± 30) K, in accordance with previous reports [49,85], $n$ = 3 [86]. The last term in relation (1) is responsible for the concave behavior, with the parameter $T_0$ = (130 ± 10) K, of the same order of magnitude as the Currie-Weiss temperature $\theta$ found in magnetic measurements (Section III.C). Consequently, $T_0$ is also the temperature scale that marks the loss of spatial correlation among magnetic moments. The concave curvature around and above $T_0$ appears as the precursor of the saturation of the electronic scattering on the maximal spin disorder in the high-temperature limit. Our interpretation is similar to the one proposed for filled cage compounds [83], where the rattling atoms, loosely bound within the crystal structure, oscillate within their cages. When fully thermally agitated, these atoms no longer contribute to the rise of configuration entropy. We also note that the temperature scale comparable to $T_0$ marks the minimum of the Seebeck coefficient in $Co_{1/3}NbS_2$. [49].

The concave curvature in $\rho_{ab}(T)$ remains preserved under pressure. Fig. 6 (a) shows excellent fits of relation (1) to experimental data of Fig. 3 (a) in the temperature range between 50 K and 300 K. The evolution of the parameters $\rho_2$ and $T_0$ under pressure is shown in Fig. 6 (b). The parameter $\rho_2$, which stands for coupling of conducting electrons to magnetic disorder, exhibits a monotonic increase under pressure. In comparison, the characteristic energy scale $T_0$ shows non-monotonic pressure dependence. Incidentally, the maximum of $T_0$ correlates with the pressure at which AF order is suppressed. Notably, the determined Debye temperature shows the expected increase under pressure, whereas the coupling $\rho_1$ [86] does not change significantly.

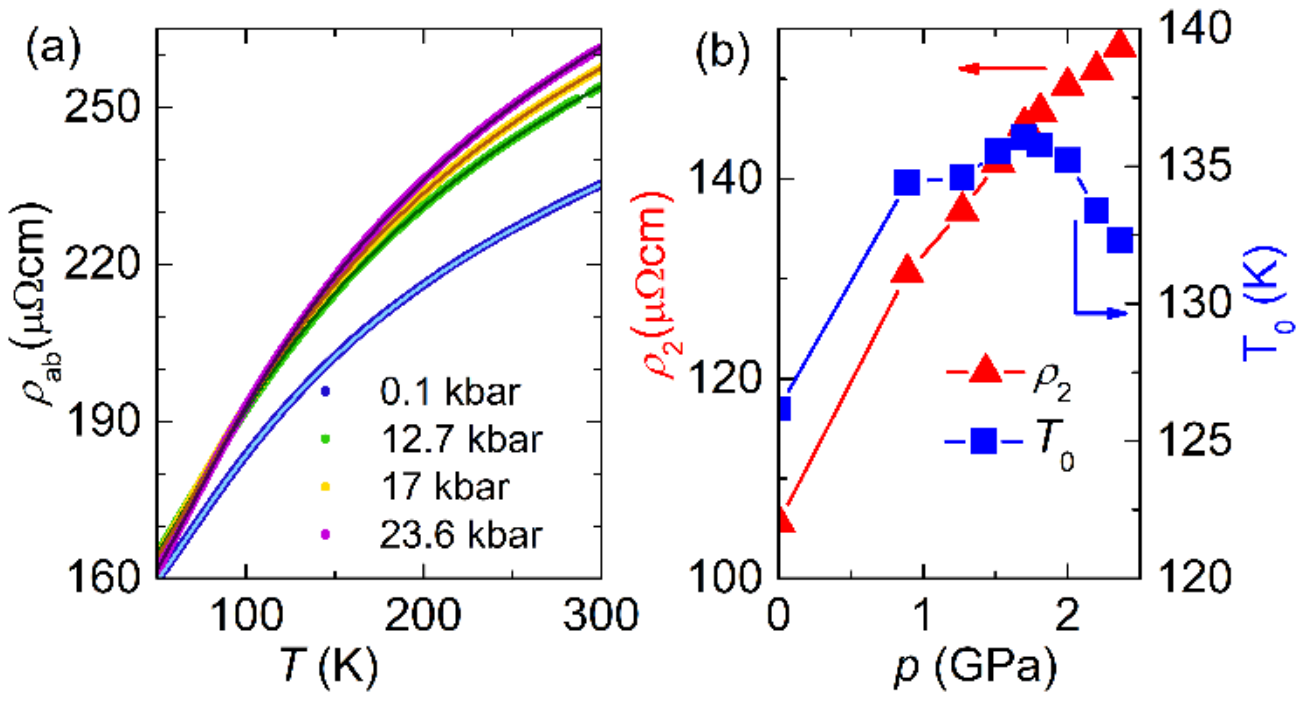

Fig. 6. Electrical resistivity at different pressures fitted above 50 K using relation (1). (b) Evolution of the $\rho_2$ and $T_0$ parameters under pressure.

Eq. (1) does not address the low-temperature features in $Co_{1/3}NbS_2$, with deviations between experimental data and fits to (1) starting to show below 50 K. Apart from the reduction of the ordering temperature under pressure, the resistivity exhibits complex temperature/pressure behavior. Several minima and maxima in $\rho_{ab}(T)$ develop below 20 K, and additional inflection points in $\rho_c(T)$. These features point to multiple low energy scales, possibly related to separate scattering mechanisms on quasi-degenerate magnetic configurations in a frustrated magnetic system [87]. The presence of complex magnetic textures $Co_{1/3}NbS_2$ has been suggested recently, even at ambient pressure, as the source of the large anomalous Hall effect [11]. Further experimental research is required to determine the evolution of spin dynamics upon varying temperature and pressure, where inelastic neutron scattering and nuclear magnetic resonance appear as promising approaches.

In contrast to the in-plane resistivity at room temperature, the resistivity along the *c*-axis decreases under pressure. The changes of opposite signs signify Co atoms' different roles in electronic transport in two directions. In both cases, the changes are probably governed by the amplification of coupling between Co orbitals and itinerant electrons of the $NbS_2$ layers. For in-plane transport, where the spin disorder on Co atoms acts primarily as the scattering source, the amplified coupling under pressure leads to higher resistivity. Conversely, in the *c*-axis direction, where Co atoms act as bridges between layers, the resistivity decreases as the coupling improves under pressure. This view also explains the big change in the residual resistivity along the *c*-axis direction under pressure. Further discussion of the electronic transport along the *c*-axis is presented in Section C, upon addressing the electronic structure.

### B. Calculations of the electronic structure in the magnetically ordered state

The simplest approach to the electronic structure of intercalates is to assume that the electronic bands of the host material remain rigid upon intercalation. The parent compound, 2H-NbS2, features fully occupied sulfur p bands and half-filled niobium *d* bands, which account for its metallic properties [88]. The bonding within layers is covalent, whereas the cohesion is much weaker between layers, often considered to originate from the Van der Waals forces. The charge transfer is expected between the intercalated ions and the conduction bands of the host, increasing their filling level. In $Co_{1/3}NbS_2$, this charge transfer has been previously argued to be one electron [40] or two electrons per Co atom [2]. The electrons left in crystal-field-split Co 3*d* orbitals are assumed to form dispersionless bands. These electrons localized on Co 3*d* orbitals are responsible for the magnetic moment of Co ions.

In reality, the number of electrons transferred from the Co atom to $NbS_2$ layers is a non-integer, whereas the hybridization between Co-orbitals and $NbS_2$ planes is finite [12] and affects the physical properties of $Co_{1/3}NbS_2$ in several ways. As we will show, the intercalation governs most of the physical processes and properties addressed in this paper, including the electric interlayer transport, the intralayer scattering of electrons, and affects the magnetic coupling between Co ions to some extent. The calculations and considerations presented in what follows explore the consequences of these hybridizations and interactions in more detail.

*Ab-initio* electronic structure computations for pristine and 2*H*-$NbS_2$ intercalated by 33% Co have been recently reviewed in Ref. [12] in conjunction with the ARPES study of the electronic structure. Most importantly, the electronic structure calculations in $Co_{1/3}NbS_2$ must account for the big magnetic moment on Co ions. This property is experimentally verified through magnetic susceptibility obeying the Currie-Weiss law in the wide temperature range above the magnetic ordering temperature. It was also verified in the magnetically ordered state through neutron scattering measurements [46]. This magnetic moment of Co ion is difficult to catch in DFT calculations in the nonmagnetic state or without properly accounting for the Coulomb repulsion within the Co *d*-orbitals. The need to address strong electron repulsion within *d*-orbitals in the *ab initio* calculations of the electronic structure of Co-based compounds is also evident from previous works [89–91], where the DFT+$U$ method has been tested and recommended. The value of $U = 5$ eV, which we obtained from DFT using the linear-response theory [64] in $Co_{1/3}NbS_2$, is consistent with those works [12]. The comparison with electronic structure calculations in the magnetic state without $U$ shows that $U$=5 eV significantly affects the distribution of electrons across Co orbitals and produces the value of Co magnetic moment closer to the one determined experimentally here. On the other hand, our investigation shows that the electronic structure close to the Fermi level is not very susceptible to small variations of $U$ around the 5 eV value.

It was shown that the DFT+$U$ calculations [12] in $Co_{1/3}NbS_2$, conducted in the magnetically ordered state, reproduce very well the most dispersive bands around the Fermi level measured in ARPES [12,50,51]. The new feature at the Fermi level observed by ARPES in $Co_{1/3}NbS_2$ that DFT+$U$ cannot catch is the shallow and weakly dispersive band presumably related to Co orbitals [12,50,51]. The observed feature, also identified in the sister compound $Cr_{1/3}NbS_2$ [92,93], is interesting from the fundamental point of view [12]. However, it is probably of secondary interest regarding the intercalation-induced effects on electric transport, where the changes in highly dispersive bands are expected to play the primary role. Thus, in this section, we concentrate on the differences between the DFT-calculated band structures of 2$H$-$NbS_2$ and $Co_{1/3}NbS_2$, shown in Fig. 7.

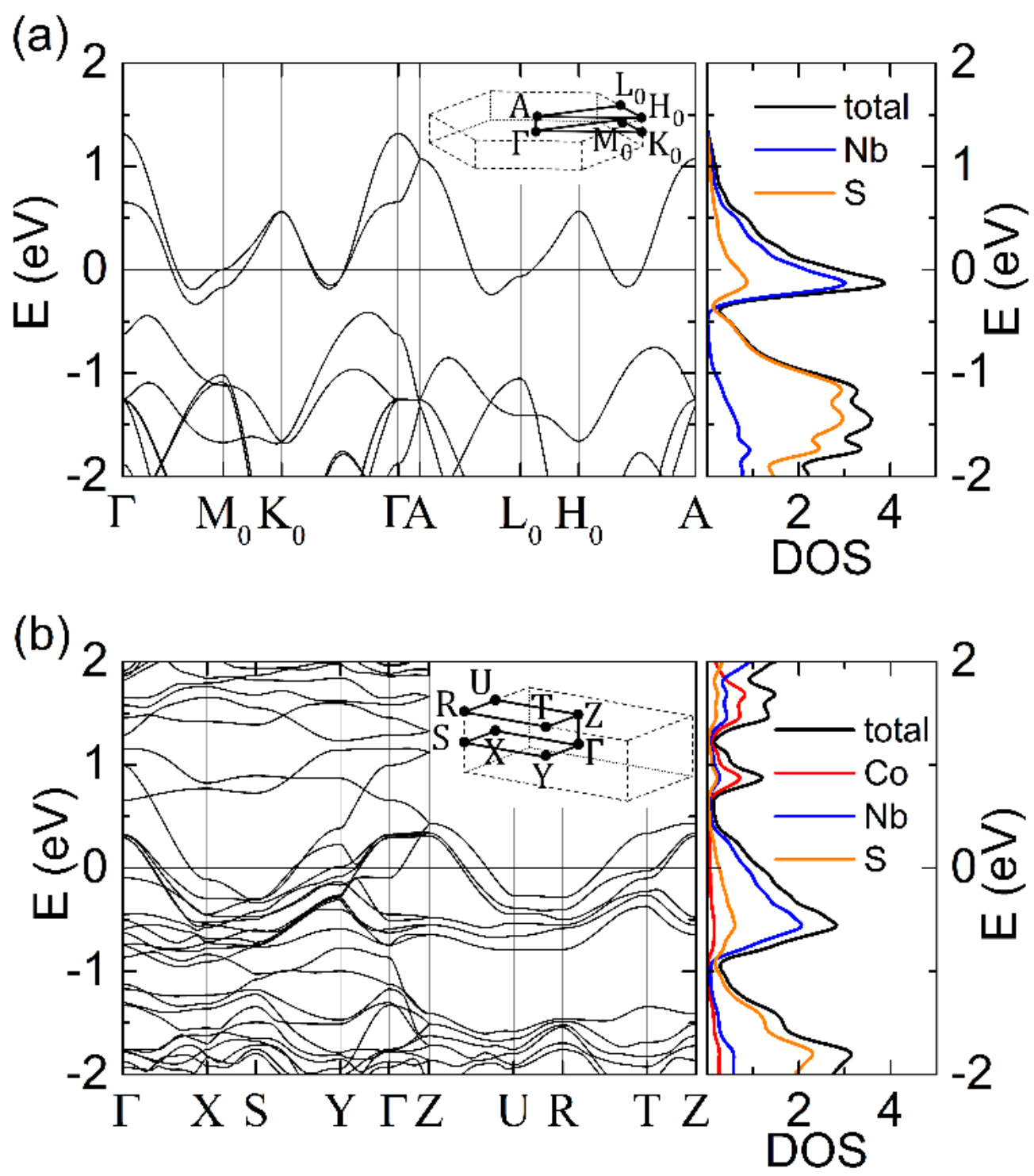

Fig. 7. Band structure for (a) 2$H$-$NbS_2$ and (b) $Co_{1/3}NbS_2$. The Fermi energy corresponds to $E = 0$.

As expected, the calculated electronic structure of $Co_{1/3}NbS_2$ inherits much from the electronic structure of 2$H$-$NbS_2$ [94]. Notably, the band structure of the single $NbS_2$ layer (not shown) is characterized by a single conduction band dominantly composed of Nb orbitals. Correspondingly, the electronic structure of the 2$H$-$NbS_2$ crystal, with two layers contributing to the unit cell, has two such bands near the Fermi level, as shown in Fig. 7(a). These bands run as quasi-degenerate throughout most of the $k$-space. The splitting between two bands, produced by the inter-layer hybridization, maximizes around Γ point, whereas no splitting occurs on the topmost surface of the first Brillouin zone (1BZ), at $k_z = \frac{\pi}{c}$. Both bands cross the Fermi level twice between Γ and K points, resulting in the Fermi surface sections shown in Fig. 8(a).

The crystallographic unit cell in $Co_{1/3}NbS_2$ is threefold bigger, with the additional doubling of the unit cell occurring upon the AF ordering. The unit cell in AF-ordered $Co_{1/3}NbS_2$ thus contains 12 Nb atoms, 24 S atoms, and 4 Co atoms. The electronic band structure of $Co_{1/3}NbS_2$ remains relatively simple near the Fermi energy, with only six bands crossing the Fermi level. The computed Fermi surfaces for both compounds can be compared in Fig. 8. Notably, the six-fold enlargement of the unit cell in real space, corresponding to a six-fold smaller first Brillouin zone in $Co_{1/3}NbS_2$ than in 2*H*-$NbS_2$, does not result in the proliferation in the number of Fermi surface segments, as one would naively expect. As discussed in more detail in Appendix A, the first reason for the relative simplicity of the calculated Fermi surface in $Co_{1/3}NbS_2$ lies in the absence of Co-dominated bands appearing at the Fermi level. The second reason lies in the shrinking of the Fermi surface pockets around Γ and $K_0$ points within the original first Brillouin zone due to increased conduction band filling. These small pockets get preserved upon folding the Fermi surface of "doped 2*H*-$NbS_2$" into the first Brillouin zone of $Co_{1/3}NbS_2$ (Appendix A).

The DFT results show the amplified separation of two original conduction bands at Γ point, indicating increased inter-plane hybridization upon Co intercalation. That leads to one of the original pockets around Γ point getting submerged below the Fermi level (see Fig. 7(a) and (b)). It is precisely due to this band evolution that the Fermi surface, in agreement with ARPES results [12], changes from a quasi-2D cylinder (Fig. 8(a)) to a more dispersive pot-like shape, central to Fig. 8(b) (details shown in Appendix A). The bottom of the "pot" represents the portion of the Fermi surface with the Fermi velocity pointing along the z-axis. In Fig. 7(b), the bottom of the pot corresponds to the band crossing the Fermi level along the ΓZ direction. Thus, we attribute the reduction of electrical resistivity along the *c*-axis and lower resistivity anisotropy to this feature of the electronic structure.

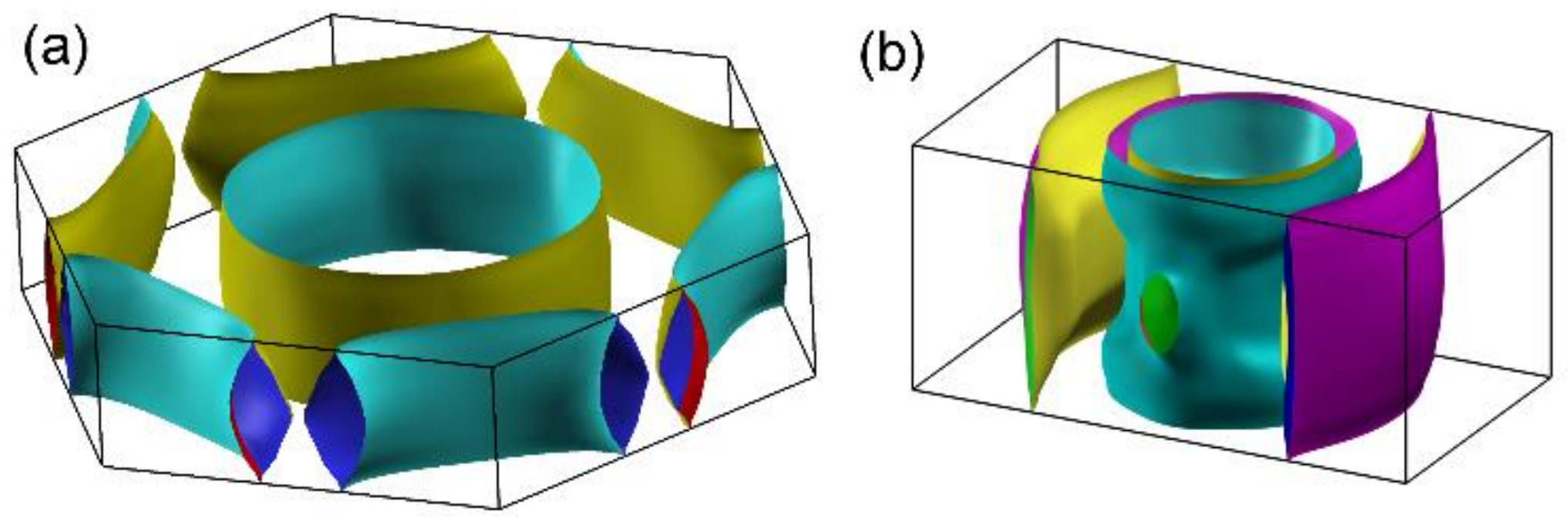

Fig. 8. Fermi surfaces of (a) 2*H*-$NbS_2$ and (b) $Co_{1/3}NbS_2$.

One way to investigate the reasons for bigger effective hybridization between Nb layers in $Co_{1/3}NbS_2$ is the following: First, the overlap between Co-orbitals and orbitals of niobium and sulfur atoms provides the covalent bonding between layers, which is not present in the parent material. This additional binding is the probable reason for reducing the interlayer distance upon intercalation, contrary to what one would usually expect to happen. Second, the reduced interlayer distance also affects the direct hybridization between $NbS_2$ layers. The relative importance of the two mechanisms is not easy to quantify. Comparing the band structure of relaxed 2*H*-$NbS_2$, the band structure of $Co_{1/3}NbS_2$, and the band structure of $NbS_2$ where the spatial arrangement of Nb and S atoms are identical as in $Co_{1/3}NbS_2$ (detailed in Appendix B) may help in that regard. The last of these band structures indicates that the second mechanism is important but insufficient to fully account for the splitting between bonding and anti-bonding Nb 4*d* bands in $Co_{1/3}NbS_2$. Further hybridization with the orbitals of the intercalated Co ions, caught by the DFT calculation in the intercalated material, is required to push the bonding band below the Fermi level. The latter hybridization also reflects in the sizable contribution of Co orbitals to the wave functions of the bonding band around the Γ point.

The contribution of various atoms and orbitals to the electronic structure can also be viewed through the projected density of states in 2*H*-$NbS_2$ and $Co_{1/3}NbS_2$, shown in Fig. 9(a)-(c).

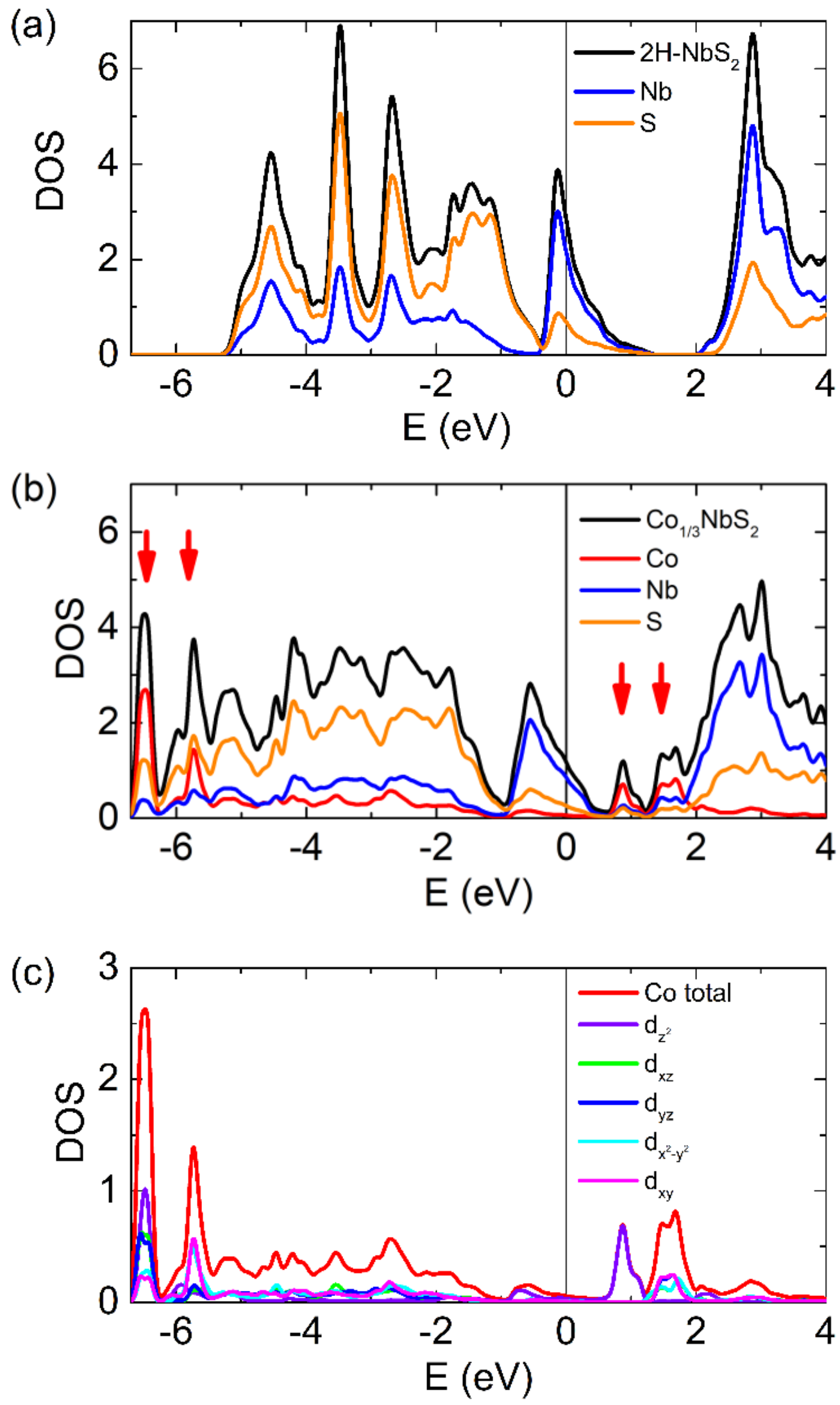

Fig. 9. The density of states (DOS) as calculated for (a) 2*H*-$NbS_2$, (b) $Co_{1/3}NbS_2$, and (c) projected on different Co-orbitals. Red arrows in (b) mark positions where most of the Co density of states is positioned. To keep the graphs for two compounds easily comparable, the units used in graphs correspond to the number of states per unit energy per unit cell divided by the number of Nb atoms within the unit cell. The origin of the energy scale is set to $E_F$.

The contribution of Co-orbitals is dominantly positioned in parts of the DOS spectra far from the Fermi level, as indicated by red arrows in Fig. 9(b). This separation primarily results from crystal field splitting and electronic correlation effects.

The part of the spectrum around the Fermi energy in Fig. 9(a) is split in half by the Fermi level, justifying the "half-filled conduction band" qualification for 2*H*-$NbS_2$. The corresponding part of the spectrum for $Co_{1/3}NbS_2$, shown in Fig. 9(b), is similar in shape but differently positioned relative to the Fermi level. The position of the Fermi level in $Co_{1/3}NbS_2$ corresponds to 5/6 filling of the "conduction-band" and seems to justify the assumed charge transfer of 2 electrons from Co ion into $NbS_2$ layers, suggested earlier [2,41,88]. However, a closer look reveals that the conduction band section of the spectrum has a bigger energy extension in $Co_{1/3}NbS_2$ than in $2H$-$NbS_2$, accompanied by a significant redistribution in the density of states. The morphing of conduction bands upon intercalation originates mainly from the bigger inter-layer hybridization, where Co orbitals contribute significantly. The participation of the Co orbitals in the conduction band leads to non-integer occupancy of the Co $3d$ orbitals. In our *ab-initio* calculations, this occupancy is determined to the value of approximately 7.5. The non-integer charge transfer also reflects in the Co magnetic moment, to be discussed in a moment.

Co-orbitals' contribution to states at the Fermi level can be visualized using the *Fermisurfer* viewer [95]. The most significant contribution from Co-orbitals appears at the bottom of the "pot-like" part of the Fermi surface (Appendix A, Fig. A2).

The local magnetic moment obtained in our DFT calculations is $\langle \mu_\alpha \rangle_{\mathrm{DFT}} = 2.36\,\mu_B$ per Co, 21% smaller than the value expected for $S = 3/2$ and $g = 2$: $\langle \mu_\alpha \rangle = 3g\mu_B S = 3\mu_B$. This reduction may result from the mixed-spin ground state, containing 36.3% of $S = 3/2$ and 63.7%. of $S = 1$ spin states on Co ions. Using those percentages to calculate the $\langle S(S+1)\rangle$ on Co ion, we obtain $\sqrt{\langle \vec{\mu}^2 \rangle} = 2\mu_B \sqrt{\langle S(S+1)\rangle}$=3.25 $\mu_B$. That is only a few percent larger than the value obtained by magnetic susceptibility measurement, $\left(\sqrt{\langle \vec{\mu}^2 \rangle}\right)_{\mathrm{exp}} = 3.17\,\mu_B$ (see Section C). It is also worth reconciling the Co magnetic moment here discussed with the charge occupancy of the Co 3d orbitals of 7.5, discussed above. Such an occupancy would imply a magnetic moment of 2.5 $\mu_B$, provided that the first Hund rule applies. This rule is indeed obeyed within our DFT results regarding the positions of the energy levels of Co $3d$ majority- and minority-spin orbitals relative to the Fermi level. The value of 2.36 $\mu_B$ is the consequence of majority-spin $3d$ orbitals on Co being *not fully occupied* due to their small participation in band states above the Fermi level.

Finally, we estimate the compressibility along the *ab* plane from the experimentally observed shifts of Bragg peaks under pressure. It amounts to 0.004 $GPa^{-1}$. The experimental setup did not permit to determine the compressibility along the c-axis. For comparison, the experimentally estimated compressibility for sister compound $Cr_{1/3}NbS_2$ amounts to 0.0022 $GPa^{-1}$ in the *ab* plane and 0.039 $GPa^{-1}$ along the *c*-axis [96]. The calculated compressibility of $Co_{1/3}NbS_2$ amounts to 0.00302 $GPa^{-1}$ in the *ab* plane and 0.00767 $GPa^{-1}$ along the *c*-axis. Thus, in magnetically intercalated TMDs, the compressibility along the *c*-axis is about twice as large as in-plane compressibility [96,97]. The ratio is much smaller than in the parent compound, where the compressibilities for two directions differ by one order of magnitude [98]. The reduced compressibility ratio indicates the transition from Van der Waals to ionic/covalent bonding between layers upon intercalation. This reduction might be viewed as the mechanical counterpart to the reduced electrical anisotropy discussed above.

### C. Spin valve regulated inter-planar electronic transport

A recent study [12] quantifies the increase in the effective hybridization between metallic Nb layers upon Co intercalation. It also indicates that this hybridization amplification is locally spin selective. Those findings are based on ARPES measurements and several types of electronic structure calculations. Here we qualitatively evoke those findings and employ them to understand the effect of magnetic ordering on the electric transport along the c-axis direction.

Here we start from the Co ion in the $3d^7$ electronic state with the +2 charge and $S = 3/2$, appropriate when Co orbitals do not hybridize with surrounding atoms but experience their crystal field. In that state, within the appropriate basis, only three Co 3*d* orbitals, approximately half-filled, participate in forming a 3/2 magnetic moment of Co. Out of three orbitals, one of them (Co $3d_z^2$) will dominantly hybridize with states forming the conduction band within the $NbS_2$ layer (that is, the Nb 4*d* orbitals and surrounding sulfur orbitals) [12], as witnessed in Fig. 9(c). This hybridization is the primary source of the observed departure of the Co ion from the ideal $S = 3/2$ state.

The local Coulomb repulsion between spin-up and spin-down electrons in Co $3d_z^2$ orbital is strong, partly represented by the *U* parameter in DFT calculations. Consequently, the Co $3d_z^2$ orbital is occupied by an electron of one spin projection (e.g., spin-down), with its energy lying well below the Fermi level. The electronic state for the electron of opposite spin projection (spin-up) within the same orbital is positioned above the Fermi level and stays empty in an ideal case. However, this last state is positioned within the 1 eV range above the Fermi level, whereas its hybridization $t_i$ with states forming the conduction band is of the order of 0.3 eV [12]. Thus, the Co atom in the spin-down state greatly increases the local hybridization between nearby Nb layers for spin-up electrons. The opposite happens for the Co atom in the spin-up state, which locally amplifies the local interlayer hybridization for spin-down electrons. The effective hybridization between Nb layers is thus significantly larger than the hybridization between Nb layers in pure 2H-$NbS_2$.

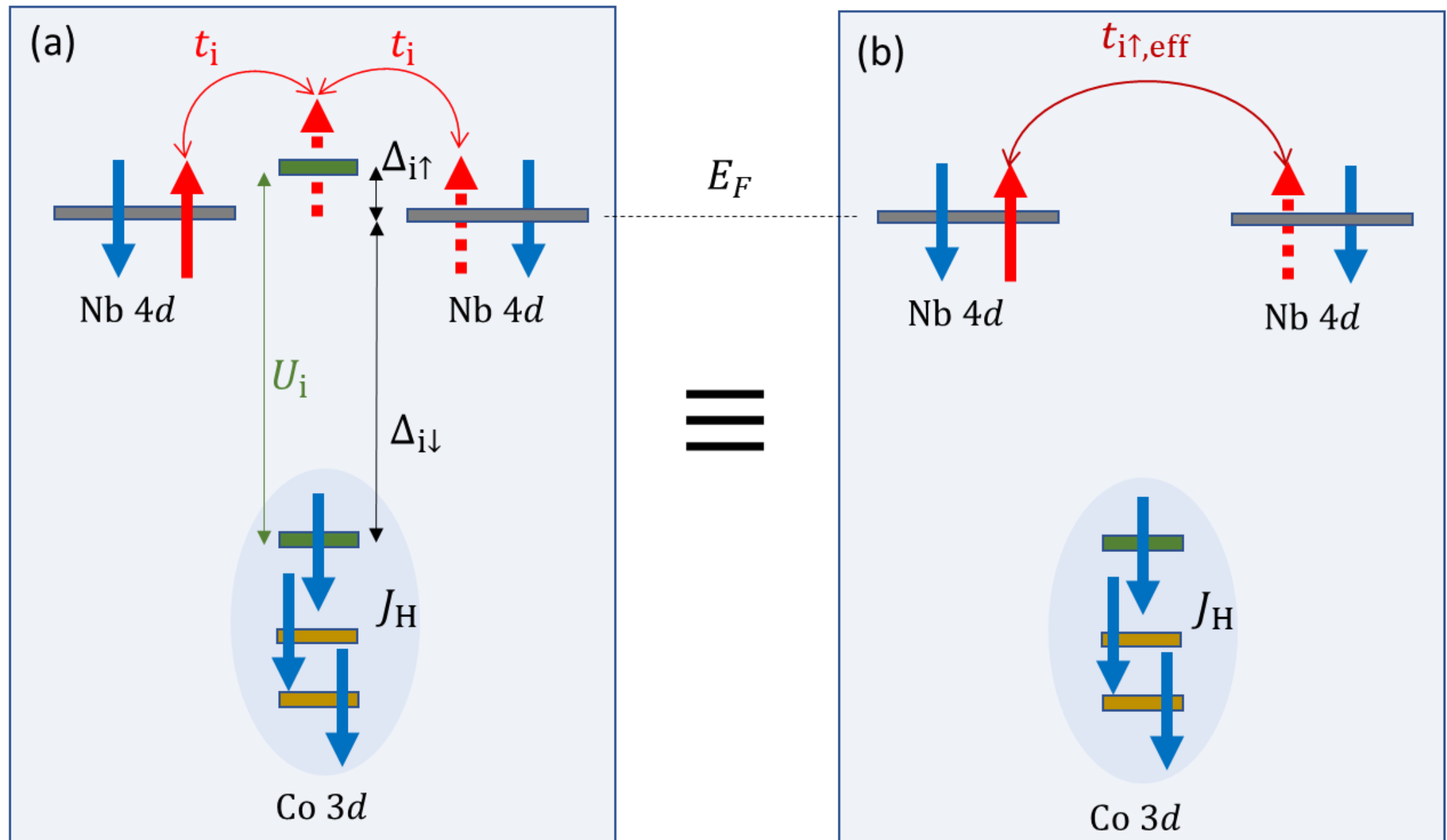

Fig. 10. The figure shows the electrons hopping between two Nb 4*d* orbitals of neighboring layers utilizing the Co $3d_z^2$ orbitals (represented by green bars) of the intercalated cobalt atom. For simplicity, only a few other Co 3*d* orbitals are shown, represented by light brown bars. The spin state of the Co atom is approximately $S = 3/2$, as ruled by the effective Hund coupling ($J_H$) between electrons Co $3d^7$ configuration. In the figure, we consider the Co atom is in the $S_z$ =-3/2 spin state. The spin-down Co 3*d* orbitals are occupied in this state, positioned deeply below the Fermi level ($E_F$), whereas the energy levels of some spin-up Co 3*d* orbitals remain empty and above the Fermi level. (a) This situation opens the possibility for spin-up electrons to hop between Nb 4*d* in neighboring layers using the unoccupied Co $3d_z^2$ spin-up orbital lying close above $E_F$. The hybridization between the Nb 4*d* orbital and the Co $3d_z^2$ orbital of the intercalated atom is denoted by $t_i$, whereas the energy separations between energy

levels of the orbitals involved are denoted by $\Delta_{i\uparrow}$ for spin-up (and $\Delta_{i\downarrow}$ for spin-down orbitals). Since $\Delta_{i\downarrow} \gg \Delta_{i\uparrow}$ for the assumed spin state of Co, the effective Co-assisted interlayer tunneling for spin-up electrons is much more efficient than for spin-down electrons, $t_{i\uparrow,\mathrm{eff}} \gg t_{i\downarrow,\mathrm{eff}}$. Only the former is shown in (b) as an effective Nb-Nb hybridization caused by the Co atom. A similar picture applies to Co in $S_z$ =+3/2 spin state, with spin-up and spin-down symbols getting interchanged. The drawing does not account for the direct Nb-Nb hybridization.

The interplanar conduction is illustrated in Fig 10. The spin-up electron (shown in red) is hopping between two Nb $4d$ orbitals in neighboring layers, virtually utilizing the Co $3d_z{}^2$ spin-up state orbital of the intercalated atom. The average spin-down state of the Co $3d_z{}^2$ orbital is maintained through the effective Hund coupling ($J_H$) to other electrons in Co $3d$ orbitals. Thus, the spin-up electrons can hop only via the spin-down Co bridges and vice versa. Equivalently, each Co ion can be regarded as a “spin-valve” regulating interlayer electronic transport. The distribution of spin-up and spin-down spin-valves depends on the magnetic state. In the antiferromagnetic HOFK state of $Co_{1/3}NbS_2$, the arrangement of Co ions obstructs the transport along the $c$-axis direction. Essentially, the distribution of “spin-valves” reduces the possibility for an electron of a given spin projection to use the nearest Co orbitals for transport between layers, forcing further in-plane “detours” instead. That effectively increases the resistivity in the magnetically ordered state compared to the magnetically disordered state, where a bigger probability exists of finding a closer appropriate bridge. The schematic drawings for the two cases are presented in Appendix C.

### D. On mechanisms of suppression of the magnetic order with pressure

Some mechanisms for suppression of the AF ordering under pressure were proposed in the previous study [49]. Here we address those scenarios in light of our new experimental findings and *ab-initio* calculations. The first scenario was motivated by the observation of strong dependence of Co magnetic moment $\mu$ on the concentration $x$ of intercalated cobalt in $Co_xNbS_2$, accompanied by the systematic variation of the $c$-axis lattice constant $c(x)$ [47]. As $x$ varies from 0.3 to 0.2, the jump in the Co spin was reported, with S diminishing from around 3/2 to below 1, accompanied by the observed decrease of the $c$-axis lattice constant. The correlation allows for the causal relationship between the value of the $c$-axis lattice constant and the Co magnetic moment. This causal relationship extends to the scenario where the Co magnetic moment gets reduced under pressure in $Co_{1/3}NbS_2$. On the other hand, our DFT+$U$ calculations for pressurized material at $x = 1/3$, show that up to hydrostatic pressure of 2 GPa, the magnetic moment of Co ion decreases less than 2% from its ambient pressure value. Thus, the DFT+$U$ calculations do not reproduce the large decrease of the Co magnetic moment, leaving the possibility that the Co magnetic moment gets heavily screened through processes not accounted for by these calculations.

The Kondo-type screening comes to mind as a possibility, alongside Doniach's mechanism for magnetic ordering collapse [99]. To recall, the Doniach phase diagram explores the competition between magnetic ordering, characterized by the ordering temperature $T_N$, and the Kondo screening, characterized by the temperature $T_K$. In Doniach's picture, where the RKKY interaction governs the magnetic ordering, $T_N$ and $T_K$ depend on the magnetic coupling $J_0$ between the local magnetic moment and conduction electrons. As the coupling $J_0$ increases, the suppression of magnetic ordering occurs when the AF-ordering temperature $T_N$, $T_N \propto J_{RKKY}$, $J_{RKKY} \propto g_F J_0^2$, becomes inferior to the Kondo temperature $T_K \propto E_F e^{-\mathrm{const.}/J_0 g_F}$, with $g_F$ denoting the electronic density of states at the Fermi level, and $E_F$ denoting the depth of the occupied metallic band. The collapse of the magnetic ordering then arises due to the pressure-induced increase of the coupling $J_0$.

There are several objections regarding the dominant role of the RKKY interaction in the magnetic ordering in $Co_{1/3}NbS_2$. The first relates to the ferromagnetic component observed in the magnetically ordered state. As discussed in Section III C, this observation indicates the presence of the Dzyaloshinskii-Moriya coupling, which is impossible to capture within the RKKY scheme. The biggest objection related to the role of the RKKY as the dominant magnetic interaction in $Co_{1/3}NbS_2$ comes from predominantly AF-character of the interlayer coupling, a benchmark of the observed HOFK magnetic state. Within RKKY, the interaction between the closest Co ions of the neighboring Co layers goes through $NbS_2$ metallic layer that they share. It is the biggest RKKY interaction in the system, and it is *ferromagnetic*. The ferromagnetic nature of the RKKY interaction at distances for $d < k_F$ is the general feature of RKKY. In our case, $d$ stands for the distance between the coupling points of two Co ions to the $NbS_2$ layer between them, and $k_F$ represents the typical Fermi wave number for the doped $NbS_2$ layer. The particularities of the electronic structure of $NbS_2$ layers and the particularities of the Co coupling to metallic states do not change that. The details of the calculations within particular models are provided in Supplementary Material.

Irrespective of the precise nature of the dominant interaction between magnetic moments, the Kondo-type screening of magnetic moments by conduction electrons remains a way to collapse magnetic ordering. The sizable reduction of the S=3/2 spin state, discussed above in relation to our *ab initio* calculations, already witnesses the presence of the strong coupling of the Co magnetic moment and conducting electrons. Addressing the Kondo-type screening in $Co_{1/3}NbS_2$ requires going beyond the DFT calculations. The recent ARPES observation of signals at the Fermi surface, not foreseen by the DFT+$U$ calculations, signals the possibility of strong electron correlations effects playing an important role in $Co_{1/3}NbS_2$ and similar magnetic intercalates [12,50,51,92,93]. It is theoretically established that S=3/2 can be effectively Kondo screened, either fully or partially ("underscreening"), depending on the number of coupled screening channels [100–103]. Spatially selective partial Kondo screening was also suggested as a possible scenario in highly frustrated Kondo lattice systems [104]. In this respect, we note that magnetic susceptibility under pressure might help to clarify remaining issues related to the Co spin state.

Finally, we come to the scenario where *magnetic frustration* is responsible for the collapse of the magnetic ordered state. The total energy calculation of various magnetically ordered states appears as a possible way to compare various magnetic configurations theoretically. This route was essentially taken by Polesya et al. [105] using DFT. However, the tripling of the unit cell upon magnetic ordering, claimed there, is not observed experimentally.

At this point, it should be stated that the DFT calculations in $Co_{1/3}NbS_2$ produce only tiny differences between total energies (per formula unit) of various magnetically ordered states. As may be expected, these differences lie within $k_B T_N \ll \theta$ range. For example, we have calculated the energies of HOFK magnetically ordered and ferromagnetically ordered states at ambient pressure. The energy of the HOFK state is lower by only 8.665 $10^{-5}$ Ry per magnetic unit cell, which roughly corresponds to 14 K. Moreover, as specified in the Methods section, these calculations require unusually large energy and wave vector cut-offs to produce consistent results. The tiny differences in energy qualitatively indicate the quasi-degeneracy of various magnetic configurations.

On the other hand, the recent study of electronic structure in $Co_{1/3}NbS_2$ [12] shows that DFT cannot reproduce certain bands observed at the Fermi level, even qualitatively. Thus our little confidence in DFT to correctly reproduce the fine energy differences between various magnetic configurations. The theoretical approaches beyond standard DFT calculations must be developed and applied to $Co_{1/3}NbS_2$ to understand its magnetic ordering from the first principles. Consequently, magnetic frustration remains a valuable candidate for understanding how the hydrostatic pressure leads to the complete suppression of the magnetic order in $Co_{1/3}NbS_2$. The quasi-degeneracy in energy between various magnetic states, being already present to a large extent at ambient pressure, can experience an additional boost under pressure. That allows for thermal and quantum fluctuations keeping the system magnetically unordered to the lowest temperatures.

## V. CONCLUSION

Our elastic neutron scattering experiments confirm the suppression of antiferromagnetic ordering in $Co_{1/3}NbS_2$ above 1.7 GPa, indicated earlier by transport measurements. For the first time in magnetically intercalated TMDs, we measure the electrical resistivity under pressure in the direction perpendicular to layers, demonstrating the unusual rise in resistivity upon entering the magnetically ordered phase. Being related to magnetic ordering, the upturn disappears as the magnetic order is suppressed by pressure. We propose spin-selective Co-assisted transport between layers as a primary reason for the observed phenomenon. Our magnetic susceptibility measurements at ambient pressure and low temperature confirm the canting of ordered Co magnetic moments, pointing to finite Dzyaloshinskii-Moriya interaction and the relevance of exchange interactions in $Co_{1/3}NbS_2$.

Several mechanisms of suppression of magnetic ordering under pressure have been explored in the paper. Concerning our new findings, some mechanisms were discarded, and some scenarios advanced and refined, with the verdict still pending. The RKKY interaction is eliminated as the dominant interaction between magnetic moments. The Kondo-type screening of magnetic moments escapes our observations but remains a candidate for the collapse of the magnetic ordering under pressure. Another viable candidate is the increased magnetic frustration under pressure, whose presence at ambient pressure is quantified through our experimental results and high cut-off DFT calculations.

Further experimental and theoretical work may be required to fully understand the nature of the magnetic state and its disappearance under pressure in $Co_{1/3}NbS_2$. The same is likely to apply to other magnetic intercalates of TMDs.

### Acknowledgments

This work was supported by the Unity through Knowledge Fund, under Grant No. 65/10, Croatian Ministry of Science, Education and Sports Grants No. 035-0352826-2848, 035-0352826-2847, and Croatian Science Foundation Projects No. IP-2016-06-7258 and IP-2020-02-9666. The work at the TU Wien was supported by FWF project P27980 - N36 and the European Research Council (ERC Consolidator Grant No 725521). The work at EPFL was supported by Sinergia grant "Mott physics beyond the Heisenberg model" of the Swiss NSF. Authors acknowledge Yuki Utsumi (Institute of Physics, Zagreb) for helpful discussions about band structure and Krešimir Molčanov (Institute Ruđer Bošković, Croatia) for determining lattice parameters of $Co_{1/3}NbS_2$.

## APPENDIX A: BAND STRUCTURE EVOLUTION THROUGH FOLDING

Fig. 7(a) in the main text shows two Nb 4*d* bands crossing the Fermi level in 2*H*-$NbS_2$ and forming Fermi surfaces shown in Fig. 8(a). In Fig. 8(b), six Fermi surfaces are shown in $Co_{1/3}NbS_2$, with one of them showing a pronounced 3D character and dispersion along $k_z$ axis. If the rigid band approximation would hold for states primarily composed of niobium and sulfur orbitals, the Fermi surface in $Co_{1/3}NbS_2$ should emerge from 2*H*-$NbS_2$ bands upon two simple transformations.

First, the Fermi level in the parent compound should be shifted to account for the charge transfer of approximately 2 electrons per Co ion into the $NbS_2$ planes. Second, the resulting Fermi surfaces should be folded from the initial hexagonal IBZ of 2*H*-$NbS_2$ into the six-fold smaller orthorhombic IBZ of AF-ordered $Co_{1/3}NbS_2$. The folding process is sketched in the $k_z = 0$ plane in Fig. A1. The folding goes in two steps. First, we fold the second and third BZ of hexagonal $Co_{1/3}NbS_2$ and obtain six cylindrical FSs: four smaller ones (full and dashed green and blue lines) and two larger ones (full and dashed orange). The second folding goes from the hexagon drawn in red in Fig. A1(b) into the twofold smaller BZ represented as the rectangle drawn in green in Fig. A1(c). As visible from Fig. A1(b) and Fig. A1 (c), only the large orange circles are affected by the last folding. The final result, shown in Fig. A1 (c), contains six Fermi surface segments arising from six bands. The folding procedure makes it easier to understand the morphing of the whole Fermi surface of 2*H*-$NbS_2$ (Fig. 8(a)) into the Fermi surface of $Co_{1/3}NbS_2$ (Fig. 8(b))

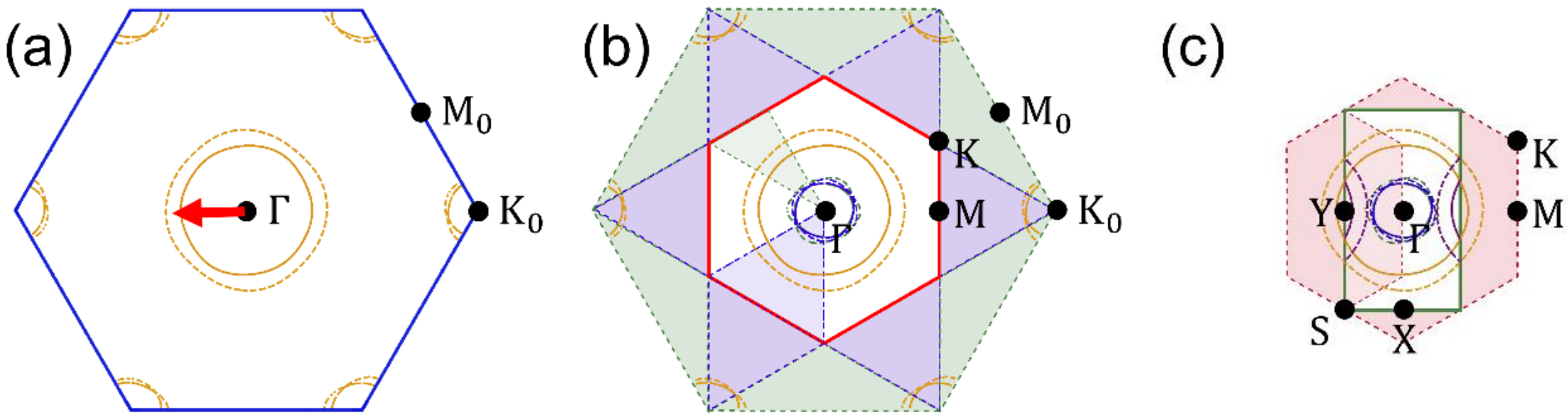

Fig. A1. Schematics of the bands (Fermi surface) folding form the large hexagonal Brillouin zone of 2*H*-$NbS_2$ to the small orthorhombic BZ of $Co_{1/3}NbS_2$ in the $k_z$=0 section: (a) Hexagonal BZ (blue) of 2*H*-$NbS_2$ with the Fermi level that accounts for the charge transfer of 2 electrons/Co from Co to $NbS_2$ planes. Two branches of FS are visible as full and dotted orange lines. (b) Hexagonal 1. BZ of $Co_{1/3}NbS_2$ (red hexagon) encompassing six Fermi surfaces: two orange circles not affected by folding, and four smaller circles - two blue circles and two green circles produced folded-in from the 2. and the 3. BZ, respectively. (c) Orthorhombic first BZ (green rectangle) of AF ordered $Co_{1/3}NbS_2$ with FS branches folded from the crystallographic first BZ (red). Four smaller FS branches are unaffected by the last folding, whereas large orange branches are partially folded. The Fermi wave number corresponding to the arrow in a) is characteristic of the almost circular Fermi surface of the uniformly doped $NbS_2$ layer.

The comparison of the folding results and the electronic structure for $Co_{1/3}NbS_2$ of Fig. 7(b) and Fig. 8(b) permits us to spot the main effect of the intercalation on the electronic dispersion: Fig. A1(c) contains *six* Fermi surface segments, arising from six bands, whereas the electronic structure results for $Co_{1/3}NbS_2$ accounts only for *five* Fermi surface segments crossing the $k_z$ =

0 plane. The corresponding five bands closely resemble the bands found in 2*H*-$NbS_2$. Fig. A2, produced through the *Fermisurfer* software [95], shows all the sections of the Fermi surface through the first Brillouin zone and resolves the mystery of the missing Fermi surface segment. The sixth band in $Co_{1/3}NbS_2$ develops much stronger dispersion in the direction perpendicular to layers than any other band in 2*H*-NbS2 or $Co_{1/3}NbS_2$ that crosses the Fermi level. The corresponding Fermi surface segment is shown in the last row of Fig A4. This "pot-shaped" segment of the Fermi surface features the significant "pot-bottom" part where the Fermi velocity points along the *c*-axis direction. Fig. A2 also provides insight into the relative importance of Co, Nb, and S orbitals in forming states at the Fermi surface. The most interesting fact to observe in Fig. A2 is that the contribution of Co orbitals to states at the Fermi surface, relatively low in general, maximizes in the "pot-bottom" part of the "pot-shaped" segment of the Fermi surface. That is also the only significant part of the Fermi surface with a substantial *z*-axis component of the Fermi velocity, likely to contribute to the electric conductivity in a direction perpendicular to layers. The high intensity of sulfur contribution in the same region suggests that the Nb-S-Co-S-Nb link provides a relatively important conducting channel for the *c*-axis electronic transport.

These observations come atop the largest share belonging to Nb2 (niobium atoms positioned right above or below the intercalated Co ions) orbitals. Their share approximately doubles the one belonging to Nb1 (niobium atoms with no Co ions above or below them in the crystal structure) orbitals, a mere consequence of the crystal structure containing twice as many Nb2 atoms than Nb1 atoms. It may also be noted that the total contribution of S orbitals is comparable to that of Nb1. In turn, the shares per atom are comparable for S and Co ions. However, the respective contributions per atom are not comparable, as S atoms are six-fold more abundant in the crystal than Nb1 atoms.

An alternative way to compare the calculated electronic structure in $Co_{1/3}NbS_2$ and 2H-$NbS_2$ was recently used in Ref. [12]. Instead of *folding-in* the 2H-$NbS_2$ spectra into the first Brillouin zone of $Co_{1/3}NbS_2$, the DFT-calculated electronic spectra of magnetically ordered $Co_{1/3}NbS_2$ were *unfolded* into the six times larger first Brillouin zone of $Co_{1/3}NbS_2$. However, the focus on interest was not the same as here, including the considerations related to electronic transport, which are one of our important concerns in the present paper.

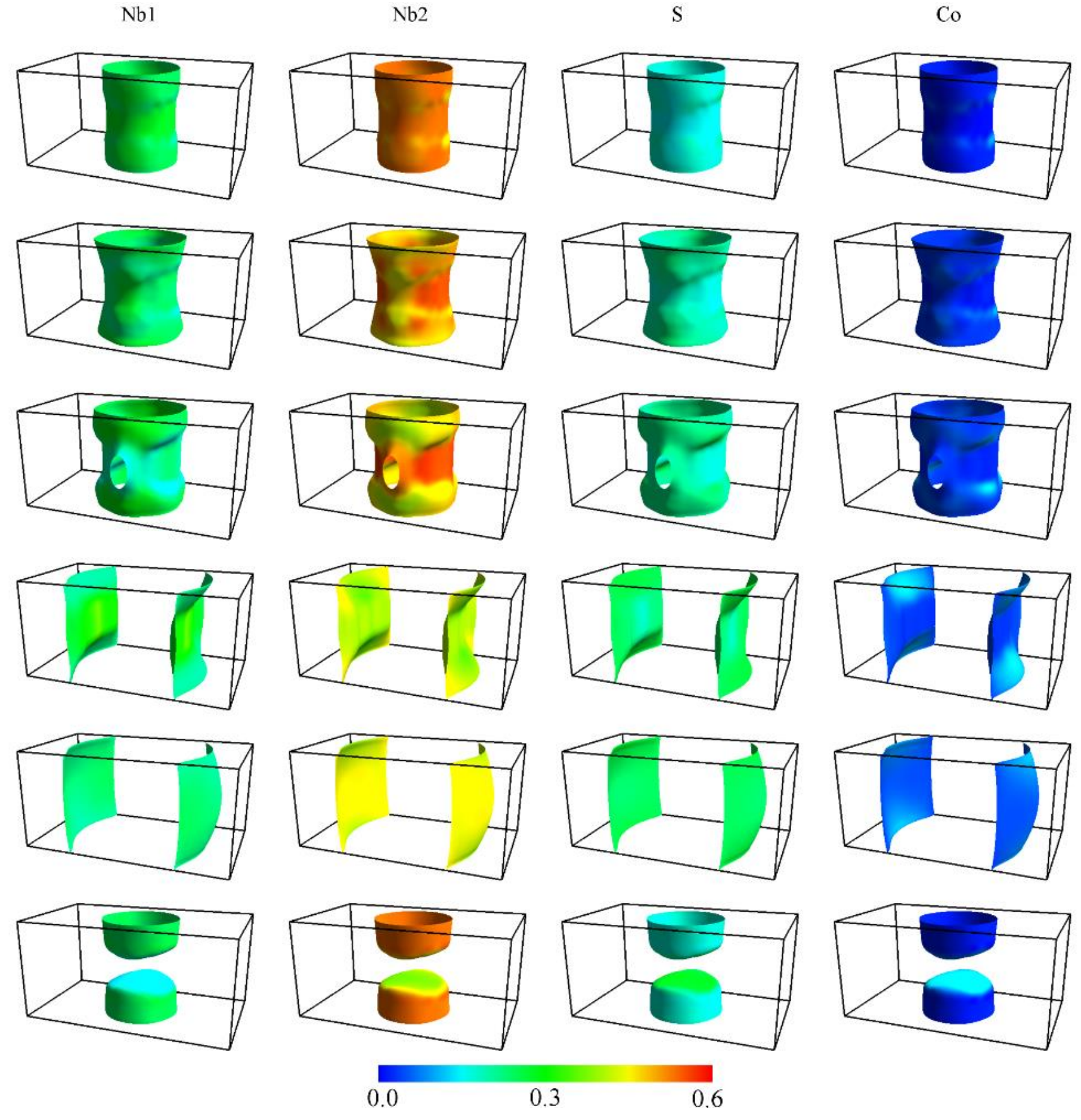

Fig. A2. The share (projection) of atomic orbitals ($d$ orbitals for Nb and Co atoms, and $p$ orbitals for S atom) in the electronic states at the Fermi surface of $Co_{1/3}NbS_2$. Nb2 stands for the niobium atoms closest to Co ions, whereas the share of other niobium atoms is denoted by Nb1.

## APPENDIX B. ELECTRONIC STRUCTURE IN "Co-DEFORMED" *2H*-$NbS_2$ CRYSTAL

To deconstruct the sources of differences in electronic structures of 2*H*-$NbS_2$ and $Co_{1/3}NbS_2$, we have calculated electronic spectra for several "auxiliary/artificial crystals." Here we show the electronic structure for the $NbS_2$ crystal, conveniently labeled "d-$NbS_2$", where niobium and sulfur atoms are positioned identically as in (DFT relaxed) $Co_{1/3}NbS_2$ crystal. The charge transfer from Co to $NbS_2$ layers is simulated by including two additional electrons per three $NbS_2$ formula units into the calculation, whereas the overall charge neutrality is maintained by adding an appropriate homogenous background charge. Regarding the $NbS_2$ planes, the difference concerning the situation experienced in $Co_{1/3}NbS_2$ is twofold: Co-orbitals are not present to hybridize with; the Coulomb potential of $Co^{2+}$ ions that strongly varies within the unit (super-)cell in $Co_{1/3}NbS_2$ is replaced by the energy offset produced by *homogeneously* distributed background charge.

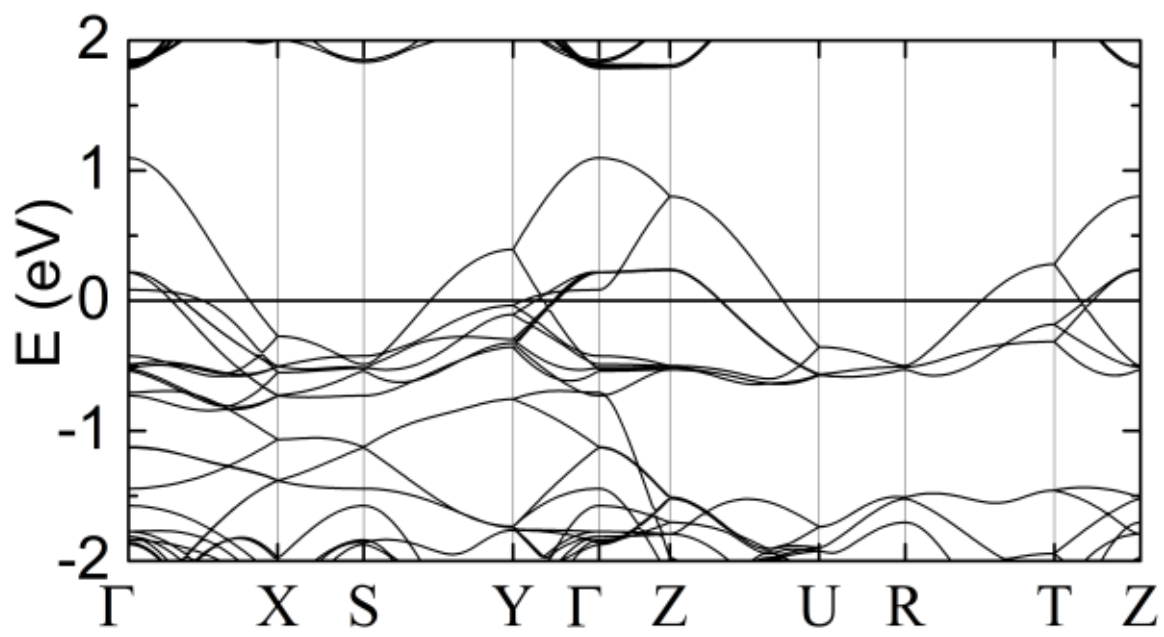

Fig. A3. The calculated electronic band structure for the "d-$NbS_2$ "crystal structure, where niobium and sulfur atoms are positioned identically as in $Co_{1/3}NbS_2$.

The results of the calculation are shown in Fig. A3. The bands that cross the Fermi level along $\Gamma - X$ line in Fig. A3 can be easily related to those appearing in $Co_{1/3}NbS_2$ and $2H$-$NbS_2$ (Section IV.B and Appendix A). The bands that meet the Γ point at 1.1 eV and 0.1 eV in Fig. A3 correspond to those already present in the same part of the Brillouin zone in $2H$-$NbS_2$ ("original-Γ" states). Their splitting at Γ point is substantially bigger than in $2H$-$NbS_2$ (Fig. 7(a)), indicating a bigger inter-layer overlap in d-$NbS_2$ than in $2H$-NbS2. The change is partly related to the smaller *c*-axis lattice constant in d-$NbS_2$. The splitting between bands increases in $Co_{1/3}NbS_2$ (see Fig. 7(b)), indicating further inter-layer hybridization occurring through Co orbitals. It should be noted that the band structure of "d-$NbS_2$" does not show the pot-like 3D Fermi surface of $Co_{1/3}NbS_2$ discussed in Appendix A.

The bundle of bands meeting the Γ point around 0.2 eV in Fig. A3 relates to the bundle of bands around the K point in $2H$-$NbS_2$ (Fig. 7(a)). In addition to being formally folded to the Γ point in $Co_{1/3}NbS_2$, these states also experience a relative shift in energy relative to "original-Γ" states. The shift is accompanied by sizable electron transfer between $K$ and Γ pockets, as first noted in the ARPES experiments in Ref. [106].

In conclusion, examining the "d-$NbS_2$" fictitious crystal helps to identify the sources of changes in the electronic structure $2H$-$NbS_2$ upon Co intercalation. It also points to the limits of the rigid-band approximation as the most straightforward approach to the electronic structure of $Co_{1/3}NbS_2$.

## APPENDIX C: SCHEMATIC C-AXIS TRANSPORT

Here we detail how the electronic transport along the *c*-axis direction is influenced by the arrangement of Co magnetic moments between layers.

Each Co ion has three closest neighbors in each adjacent layer. Out of those three neighbors in each layer, in the HOFK state [46], one is in a *ferro*-arrangement, and two have opposite magnetic moment orientation from the central one. If we consider the central Co ion to be in a "spin-up" state, then, according to Fig. 10 of the main text, only spin-down electrons will benefit from Co-assisted transport. Now, when the spin-down electron arrives in the next layer, to continue along the *c*-axis, it needs to find its following spin-up Co ion. However, as mentioned above, it can use only one of the three closest Co ions or make a larger detour in the plane. That suppresses the transport along the c-axis in the ordered state compared to the disordered state where the probabilities of finding Co ion in each spin state are equal. The situation is schematically illustrated in Fig A4.

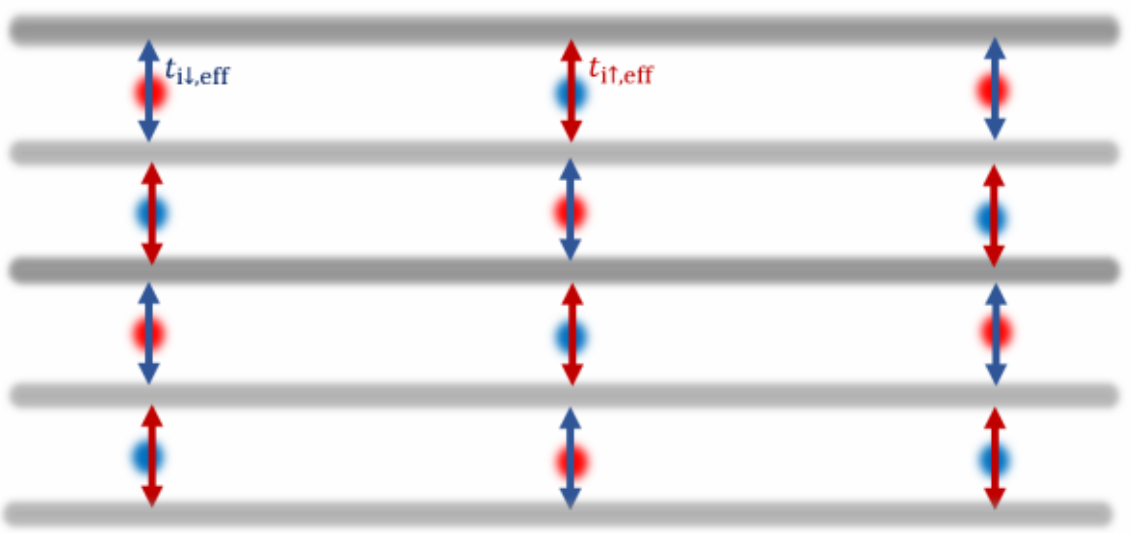

Fig. A4 The schematic representation of dominant interlayer hybridization in the magnetically ordered state. The $NbS_2$ metallic layers are shown as thick horizontal grey lines with soft edges. The red and soft blue balls represent Co ions in spin-up and spin-down magnetic states. The double arrows in red and blue stand for locally enhanced hybridizations in the spin-up and spin-down channels, respectively. Of course, the 2D representation, made here for simplicity, cannot faithfully reflect the 3D atomic structure of $Co_{1/3}NbS_2$, nor its HOFK magnetic state. The essential features captured by the drawing are the average antiferromagnetic characters of the inter-layer and the intra-layer orderings.

Fig. A5 qualitatively explains why the c-axis resistivity in the material increases upon ordering or decreases as the magnetic ordering loses coherence. In essence, for electrons of any spin projection, the AF-ordering tends to misalign the Co-created tunneling bridges along the c-axis direction. In HOFK magnetically ordered state, the electron of any spin projection is obliged to travel longer sections within planes before taking advantage of Co-created interlayer tunneling bridges (Fig. A5 (a)). The disruptions of the magnetically ordered state bring in the shortcuts along the c-axis direction (Fig. A5 (b)), leading to shorter effective paths and lower resistivity.

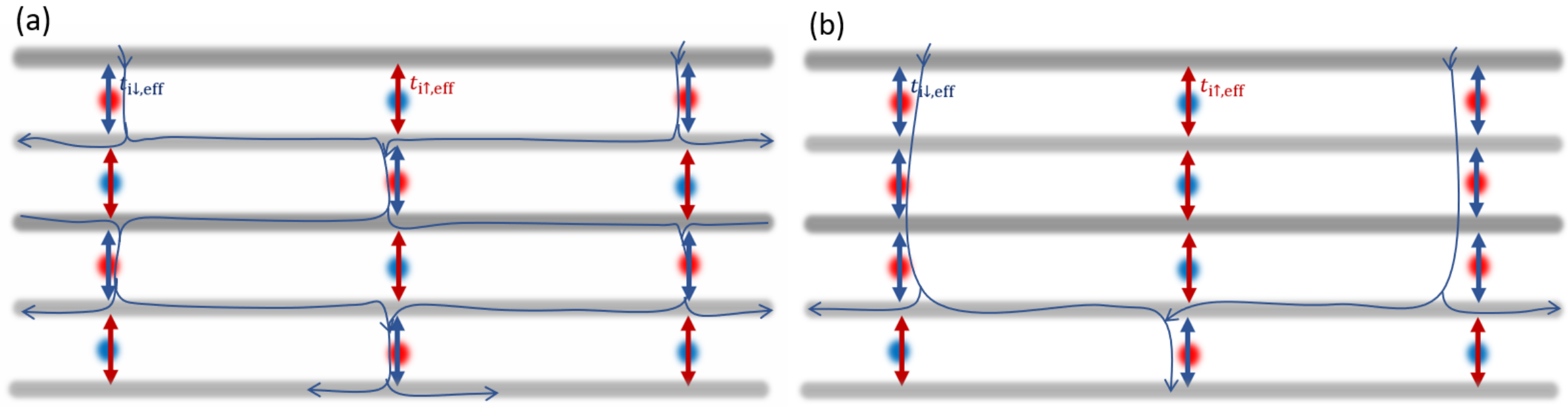

Fig. A5. The conduction paths that (spin-down) electrons take in the electric field perpendicular to layers: (a) in the magnetically ordered state in $Co_{1/3}NbS_2$ and (b) in the state where the magnetic order is locally disrupted. For example, we flip the magnetic moments at three Co ions in the second layer from the top. The meanings of symbols are the same as in Fig. A4. The paths of electrons are shown in curved blue lines. The arrows indicate the direction of electron propagation carrying the current along the c-axis direction. In the magnetically ordered state, the electron must travel larger in-plane sections before taking advantage of a suitable interlayer tunneling bridge provided by a Co-ion. A local disruption in magnetic order along the c-axis direction provides shorter paths with less in-plane scattering.

Supplemental Material to

# Electronic transport and magnetism in the alternating stack of metallic and highly frustrated magnetic layers in $Co_{1/3}NbS_2$

P. Popčević, I. Batistić, A. Smontara, K. Velebit, J. Jaćimović, I. Živković, N. Tsyrulin, J. Piatek, H. Berger, A. Sidorenko, H. Rønnow, L. Forró, N. Barišić, E. Tutiš

## On the RKKY interaction between Co magnetic ions in $Co_{1/3}NbS_2$

## Abstract

This note addresses whether the RKKY interaction can be regarded as the most relevant interaction in $Co_{1/3}NbS_2$, responsible for the magnetic ordering of Co ions, as observed in the compound. We show that this cannot be the case. The perturbative approach that underlies the RKKY interaction demands that the coupling of Co ions to $NbS_2$ layers can be regarded as weak. In the paper's main text, we have already shown that this basic assumption is wrong, as large spin-dependent Co-mediated hybridization between $NbS_2$ layers is required to produce the observed splitting between bonding and the antibonding conduction bands. This large splitting is observed in ARPES and reproduced by the *ab-initio* electronic calculations. Despite this main objection, we consider the question of RKKY in $Co_{1/3}NbS_2$ of some academic interest. Thus, we pursue the "RKKY in $Co_{1/3}NbS_2$" exercise to some length here. We start from the *rigid* metallic bands of the $NbS_2$ layers. Appropriate for the rigid-band picture of the intercalate, these bands are assumed to host two extra electrons per three $NbS_2$ formula units transferred from the intercalated Co ions, making them 5/6 filled. We concentrate on the sign of the Co-interlayer coupling. The question is addressed through a series of calculations, where the model complexity increases at each stage. The purpose of this procedure is to disclose the reasons behind the result instead of just stating the result of numerical calculations. The main result is that the dominant interaction between successive Co sublayers is *ferromagnetic* within the RKYY calculation, opposing the antiferromagnetic nature of the observed magnetic order.

## 1. Introduction

The RKKY mechanism is the magnetic interaction between magnetic ions mediated by metallic subsystem based on partially filled (conduction) bands (Ruderman and Kittel 1954; Kasuya 1956; Yosida 1957). The mechanism accounts for one magnetic ion causing the spin-polarization of the metallic electron gas, further experienced by another magnetic ion somewhere else. The RKKY interaction is primarily determined by the spin susceptibility, $\chi_s(\vec{q})$ of the metallic gas and its Fourier transform to direct space, $\chi_s(\vec{R})$. The local interaction $-J_0\overrightarrow{M_1}\cdot\vec{s}$ between magnetic ions and the spin of the metallic system sets the spin polarization response to $\vec{s}(\vec{R}) = \chi_s(\vec{R})J_0\overrightarrow{M_1}$, and the overall interaction $J_{RKKY}(\overrightarrow{R_1},\overrightarrow{R_2})= -J_0^2\chi_s(\vec{R})\overrightarrow{M_1}\cdot\overrightarrow{M_2}$. In general, $J_{RKKY}(\overrightarrow{R_2} - \overrightarrow{R_1})$ shows the power law decay and oscillates in sign as the function of distance, $J_{RKKY}(\overrightarrow{R_2} - \overrightarrow{R_1})$ is invariably *ferromagnetic* at very short distances. Formally, for $\overrightarrow{R_1} = \overrightarrow{R_2}$ the second magnetic moment experiences the same magnetic surroundings as the first one, which has polarized the electronic medium to reduce the coupling energy. How far in space this ferromagnetic region extends depends on the precise electronic structure of the metal and its Fermi wave-vector in particular.

## 2 Coupling of Co magnetic moment to NbS2 layers

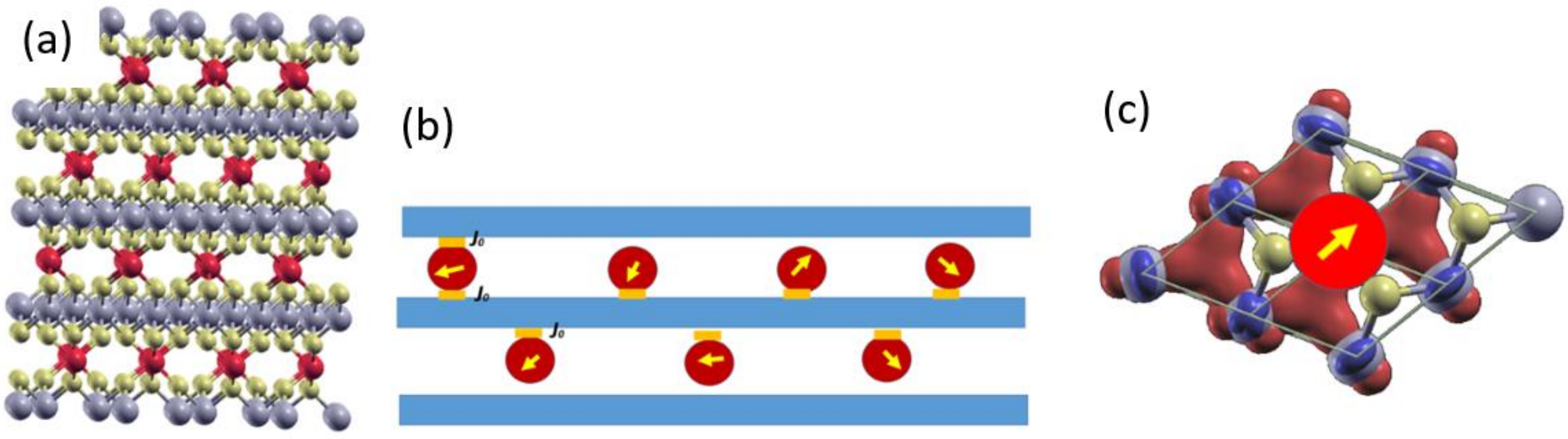

Fig. S1: The coupling of the Co magnetic moments to $NbS_2$ layers (a): The 3D representation of the crystal structure of $Co_{1/3}NbS_2$, with Co, Nb, and S atoms shown in red, grey, and yellow, respectively. (b) The simplified representation in 2D of the Co magnetic moments coupled to metallic layers is shown in blue. The magnetic coupling $J_0$ to the middle layer is graphically represented through orange rectangles. The main point is that the dominant coupling between Co magnetic moments in subsequent Co sublayers occurs through a single $NbS_2$ layer between Co sublayers. Note also that the nearest-neighboring Co couplings belong to Co atoms of subsequent Co-sublayers. (c) The Co magnetic orbital is viewed from the direction perpendicular to layers. The Co orbital is positioned above the junction of three Nb-dominated orbitals (shown in darker red), forming the conduction band with the $NbS_2$ layer. These orbitals represent the maximally localized wave-functions emerging from the Wanier90 procedure. The geometry of hybridization between Co and Nb orbitals sets the form of the magnetic coupling between Co magnetic moments and the conduction bands.

Our calculation of the coupling matrix elements between various Co sites and $NbS_2$ layers is based on our DFT calculations in $Co_{1/3}NbS_2$. The host electronic spectrum is based on our DFT calculations in 2H-$NbS_2$. Both calculations are further processed through Wannier90 calculation. The conduction bands are thus described by maximally localized Wannier functions, which hybridize through the effective tight-binding scheme and faithfully reproduce the dispersion obtained from the Quantum Espresso DFT calculation. The $NbS_2$ electronic model is used in the RKKY calculation, whereas the form of the coupling between Co and Nb-orbitals is taken from the $Co_{1/3}NbS_2$ model (Popčević et al. 2022). The latter is illustrated in Fig. S1.

## 3. Electronic band structure of the host compound

The electronic structure of 2H-$NbS_2$ is discussed in some detail in the paper's main text. The Fermi surface of 2H-$NbS_2$ is strongly two-dimensional. The electronic dispersion in the direction perpendicular to layers is weak, measured by the effective-interlayer hybridization integral $t_\perp$ being much smaller than the in-plane electronic bandwidth, $W_\parallel$ , $|t_\perp / W_\parallel| \ll 1$. Appropriate for the rigid-band picture of the intercalate, the $NbS_2$ rigid bands gain two extra electrons per three $NbS_2$ formula units, transferred from the intercalated Co ions. The band filling is thus 5/6. The dispersion within $NbS_2$ planes is predominantly characterized by circular Fermi surfaces of the quasi-2D electron gas, as shown in Fig. S2.

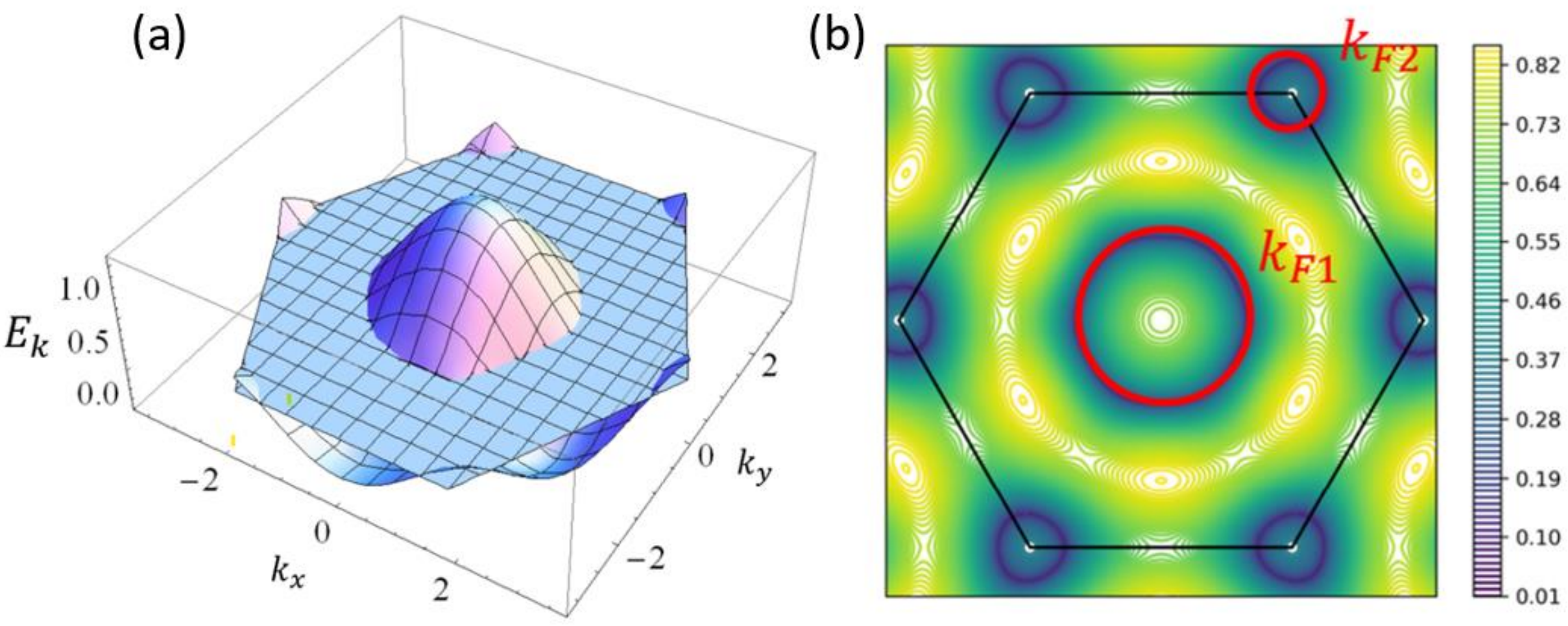

Fig. S2: The in-layer electronic dispersion used in the calculation. (a) The plot of the electronic dispersion, with the horizontal planar surface representing the Fermi level. The cuts between two surfaces mark the 2D Fermi surface at 5/6 band filling. (b) The contour plot of $(E_k - E_F)^2$. The black regions represent the 2D Fermi surface. The red circles mark two types of hole pockets being found around the central part of the Brillouin zone, and its corners, with respective radii denoted by $k_{F1}$ and $k_{F2}$.

## 4. RKKY interaction within the 2D jellium model for $NbS_2$ layers

The analytical solution for the RKKY interaction is possible under the assumption that one of the two types of Fermi pockets with circular Fermi surfaces, characteristic of 2H-$NbS_2$, can be assumed to influence the RKKY interaction dominantly. The expression for the magnetic response of the 2D electron gas to the localized disturbance can be found in Ref. (Béal-Monod 1987). The response in direct space is given by the susceptibility $\chi_{s,2D}^{0}(r)$ given by the formula,

$$\chi_{s,2D}^{0}(r) = -g_F k_F^2 [J_0(k_F r) Y_0(k_F r) + J_1(k_F r) Y_1(k_F r)] \,. \qquad (1)$$

Here $r$ marks the distance from the localized disturbance, whereas $g_F$ and $k_F$ stand for the density of states and the Fermi wave vector of the electronic system, respectively. Symbols $J_0, J_1$ and $Y_0, Y_1$ are the usual ones for the Bessel functions of the first and the second kind.

The susceptibility$\chi_{s,2D}^{\ 0}(r)$ shows a power-law decrease with distance, with the alternations in sign over characteristic distance set by the Fermi wave-vector $k_F$ of the electron gas. The results for the RKKY couplings in the jellium model are shown in Fig. S3. The figure presents the result for $k_F = k_{F1}$, corresponding to the hole pocket around the center of the Brillouin zone in Fig. S2. The most important detail to observe here is the sign of the coupling $J_1'$ between closest Co atoms in neighboring Co sublayers. This coupling turns out *ferromagnetic*. As stated already, the RKKY interaction always turns ferromagnetic at short distances, $d < 1/k_F$. Since $1/k_{F1} < 1/k_{F2}$, the same ferromagnetic nature of the $J_1'$ coupling also applies for the jellium model with $k_F = k_{F2}$, corresponding to the hole pockets around the center corner of the Brillouin zone. The ferromagnetic nature of the $J_1'$ coupling is inconsistent with the type of magnetic order found in $Co_{1/3}NbS_2$.

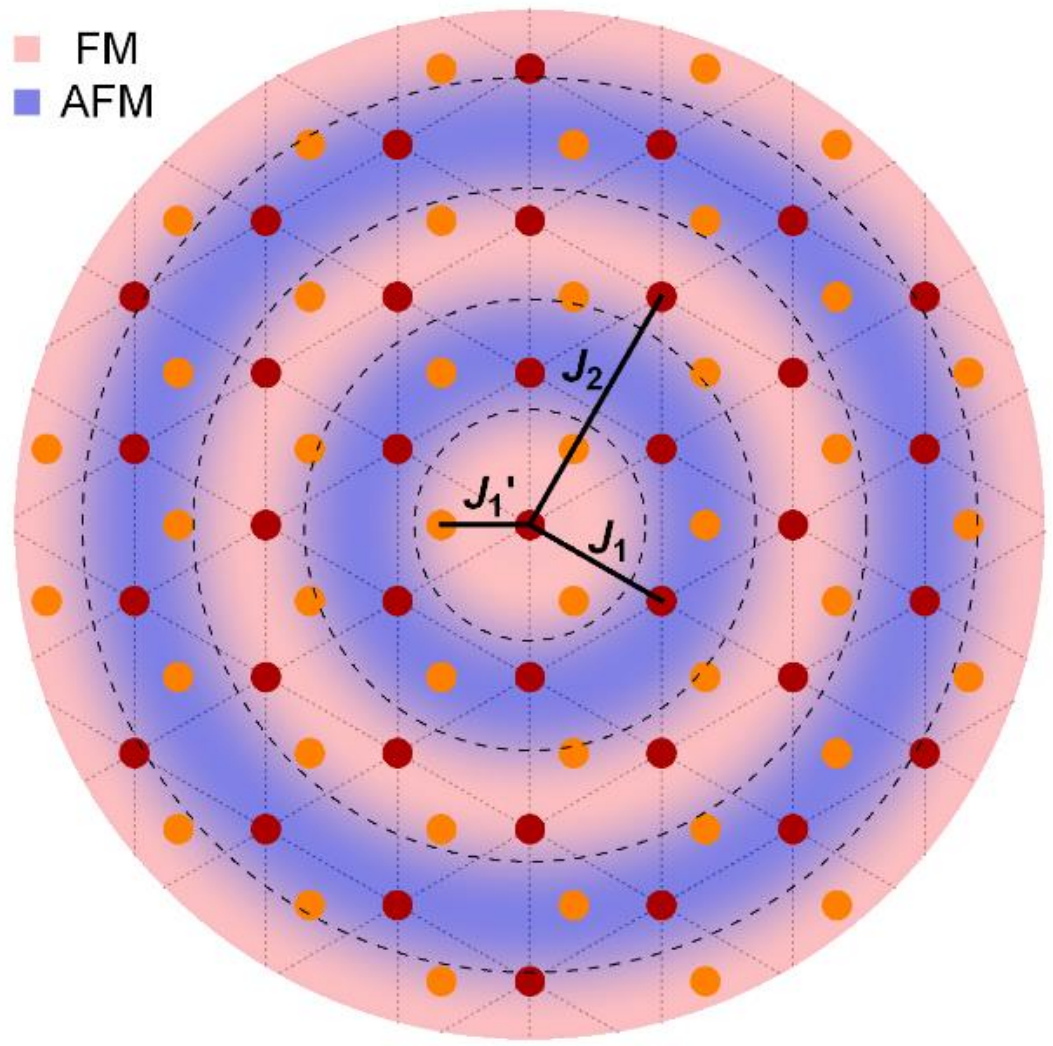

Fig. S3: The spatial variation of the RKKY-part of magnetic interaction between the Co-atom placed in the center and other Co atoms in the same Co-sublayer (brown dots) and the adjacent Co-sublayers (orange dots). The variation follows Eq. (1), with the asymptotic $r^{-2}$ decay in amplitude is compensated for to emphasize the variation in sign. $k_F = k_{F1}$, corresponding to the hole pocket around the center of the Brillouin zone in Fig. S2. The light and dark areas correspond to ferromagnetic and antiferromagnetic interaction, whereas dashed circles mark the radii where the interaction changes sign.

## 5. Effects of small interlayer hybridization

The RKKY interaction between Co magnetic moments, coupled to the same metallic layer, from its both sides, is dominated by the in-layer electronic dispersion, shown in Fig. S2. The effect of the interlayer hybridization $t_\perp$ is expected to provide a minor correction. The exact solution for a quasi-2D system with weak interlayer hybridization and for homogeneous electron gas within layers can be found in (Aristov 1997). Applied to our case of Co-intercalated 2H-$NbS_2$, the corrections of the order of $(t_\perp / W_\parallel)^2$ appear, leading to $J_x^{(3D)} = J_x^{(2D)} \cdot \left(1 - O\left(t_\perp^2/W_\parallel^2\right)\right)$, where $J_x$ denotes any of particular magnetic couplings.

## 6. The RKKY within the 2D tight-binding model for the $NbS_2$ layers

Departing from the model of the homogeneous jellium model of electrons within $NbS_2$ layers, we now account for a more realistic electronic dispersion of the conduction band of the $NbS_2$ layer shown in Fig. S2. The fact that the Fermi surface has more segments than one, and more characteristic wave-vectors than one, leads to some interference effect in the magnetic susceptibility $\chi(q)$ in the inverse space. Various contributions can be appreciated from Fig. S4, shown below.

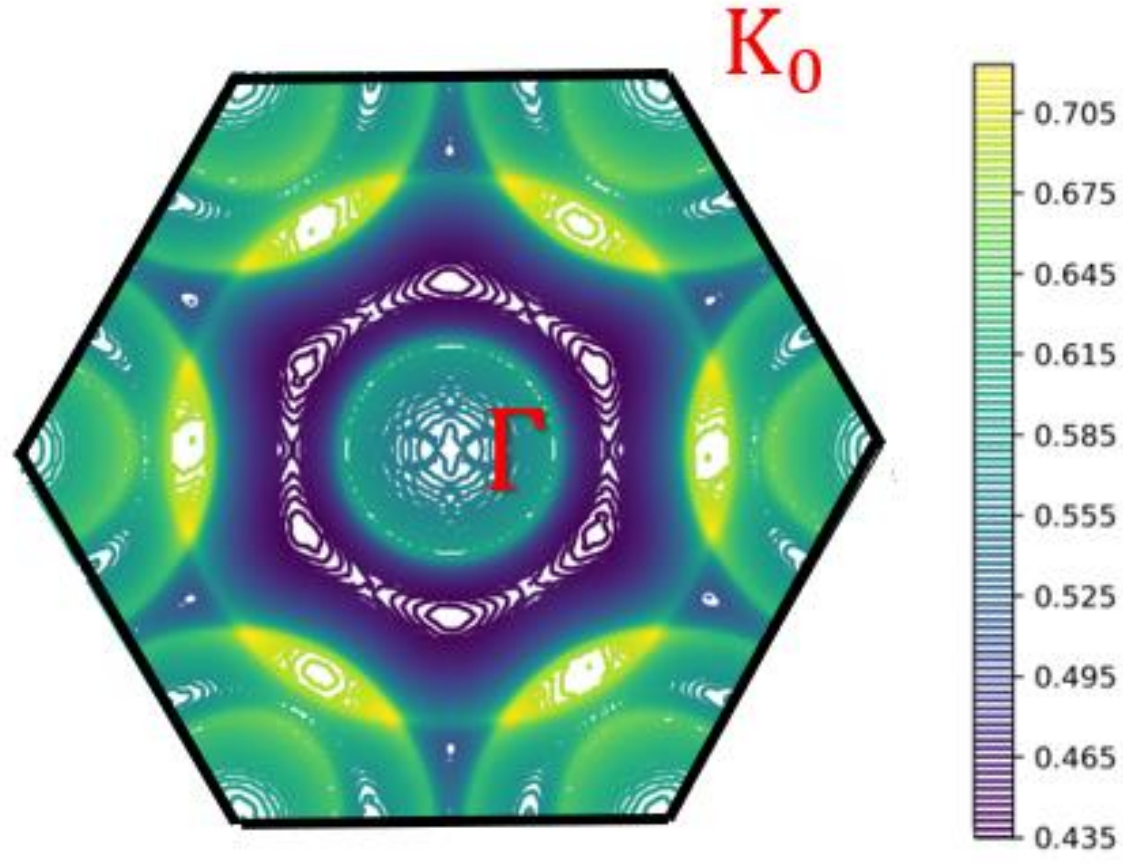

Fig. S4: The susceptibility $\chi(q)$ resulting from the electronic dispersion shown in Fig. S2. Apart from the usual signals at $|q| = 2k_{F1}$ and $|q| = 2k_{F2}$, the other signals appear as results of interferences. However, the strongest signal within the first Brillouin zone occurs at $|q| = 2k_{F1}$, indicating a possible dominant length scale of $1/2k_{F1}$, as in the case leading to Fig. S3.

The RKKY coupling constants in direct space can be calculated through the numerical Fourier transformation of $\chi_s(q)$, or directly through the formula (Suzuki et al. 1989)

$$J(\boldsymbol{R_1} - \boldsymbol{R_2}) = J_0^2 2 \sum_{\boldsymbol{k},\boldsymbol{k}'} f(\varepsilon_{\boldsymbol{k}}) \frac{M(\boldsymbol{r_1},\boldsymbol{r_2})}{\varepsilon_{\boldsymbol{k}'} - \varepsilon_{\boldsymbol{k}}} \cos[(\boldsymbol{k}' - \boldsymbol{k})(\boldsymbol{l}_1 - \boldsymbol{l}_2)] \qquad (2)$$

It turns out that the *ferromagnetic* Co-interlayer coupling $J_1{}'$ persists as the result.